\documentclass[twocolumn,prb,aps,reprint,floatfix,nofootinbib,longbibliography,superscriptaddress]{revtex4-2}
\usepackage[colorlinks=True,citecolor=Blue,linkcolor=BrickRed,urlcolor=Blue]{hyperref}
\usepackage{amsmath}
\usepackage{amssymb}
\usepackage{amsthm}
\usepackage[usenames,dvipsnames]{xcolor}
\usepackage{braket}
\usepackage{graphicx}
\usepackage{dsfont}
\usepackage{soul}
\usepackage{float}

\def \mr{\mathrm}
\def \mc{\mathcal}

\def \mb{\mathbf}

\def \d{\mathrm{d}}
\def \io{\mathrm{i}}
\def \zcap{\hat{e}_{z}}

\def \dS{\delta \mathbf{S}}
\def \bs{\boldsymbol}

\newcommand {\apgt} {\ {\raise-.5ex\hbox{$\buildrel>\over\sim$}}\ }
\newcommand {\aplt} {\ {\raise-.5ex\hbox{$\buildrel<\over\sim$}}\ }

\newcommand {\rem}[1]{}

\def  \w{\omega}

\def  \G{\Gamma}
\def  \q0{\frac{\w_k^2}{\G}+z}

\def \intk1k2{\int_{-\pi}^{\pi}\frac{\d k_1}{2\pi} \frac{\d k_2}{2\pi}}
\def \intq1{\int_{-\pi}^{\pi} \frac{\d q_1}{2\pi}}

\begin{document}
\title{How many-body chaos emerges in the presence of quasiparticles}
\author{Sibaram Ruidas}
\email{sibaram.ruidas@icts.res.in}
\affiliation{International Centre for Theoretical Sciences, Tata Institute of Fundamental Research, Bengaluru 560089, India}
\author{Sthitadhi Roy}
\email{sthitadhi.roy@icts.res.in}
\affiliation{International Centre for Theoretical Sciences, Tata Institute of Fundamental Research, Bengaluru 560089, India}
\affiliation{Max-Planck-Institut f\"{u}r Physik komplexer Systeme, N\"{o}thnitzer Stra{\ss}e 38, 01187 Dresden, Germany}
\author{Subhro Bhattacharjee}
\email{subhro@icts.res.in}
\affiliation{International Centre for Theoretical Sciences, Tata Institute of Fundamental Research, Bengaluru 560089, India}
\affiliation{Max-Planck-Institut f\"{u}r Physik komplexer Systeme, N\"{o}thnitzer Stra{\ss}e 38, 01187 Dresden, Germany}
\author{Roderich Moessner}
\email{moessner@pks.mpg.de}
\affiliation{Max-Planck-Institut f\"{u}r Physik komplexer Systeme, N\"{o}thnitzer Stra{\ss}e 38, 01187 Dresden, Germany}

\begin{abstract}
Many-body chaos is a default property of many-body systems; at the same time, near-integrable behaviour due to weakly interacting quasiparticles is ubiquitous throughout condensed matter at low temperature. There must therefore be a, possibly generic, crossover between these very different regimes. Here, we develop a theory encapsulating the notion of a cascade of lightcones seeded by sequences of scattering of weakly interacting harmonic modes as witnessed by a suitably defined chaos diagnostic (classical decorrelator) that measures the spatiotemporal profile of many-body chaos. Our numerics deals with the concrete case of a classical Heisenberg chain, for either sign of the interaction, at low temperatures where the short-time dynamics are well captured in terms of non-interacting spin waves. To model low-temperature dynamics, we use ensembles of initial states with randomly embedded point defects in an otherwise ordered background, which provides a controlled setting for studying the scattering events. The decorrelator exhibits a short-time integrable regime followed by an intermediate `scarred' regime of the cascade of lightcones in progress; these then overlap, leading to an avalanche of scattering events which finally yields the standard long-time signature of many-body chaos. 
\end{abstract}

\maketitle

\section{Introduction \label{sec:intro}}

How does chaos emerge from the dynamics of a many-body system? 
While simply stated, this question is fundamental and rather subtle~\cite{pikovsky1994roughening}. 
This is because the universal, low-temperature properties of generic condensed matter systems are not explicitly chaotic.
Instead, they are often thought of as being essentially integrable for a large class of systems: the dynamics is governed by low-energy excitations of a ground state which are well described by weakly interacting quasiparticles~\cite{chaikin2000principles}. 
And yet, at late times, the dynamics is chaotic. 
The search for a, possibly generic, mechanism underpinning the crossover from low-temperature near-integrability to late-time chaos constitutes the central objective of this work.

\begin{figure}
\includegraphics[width=\linewidth]{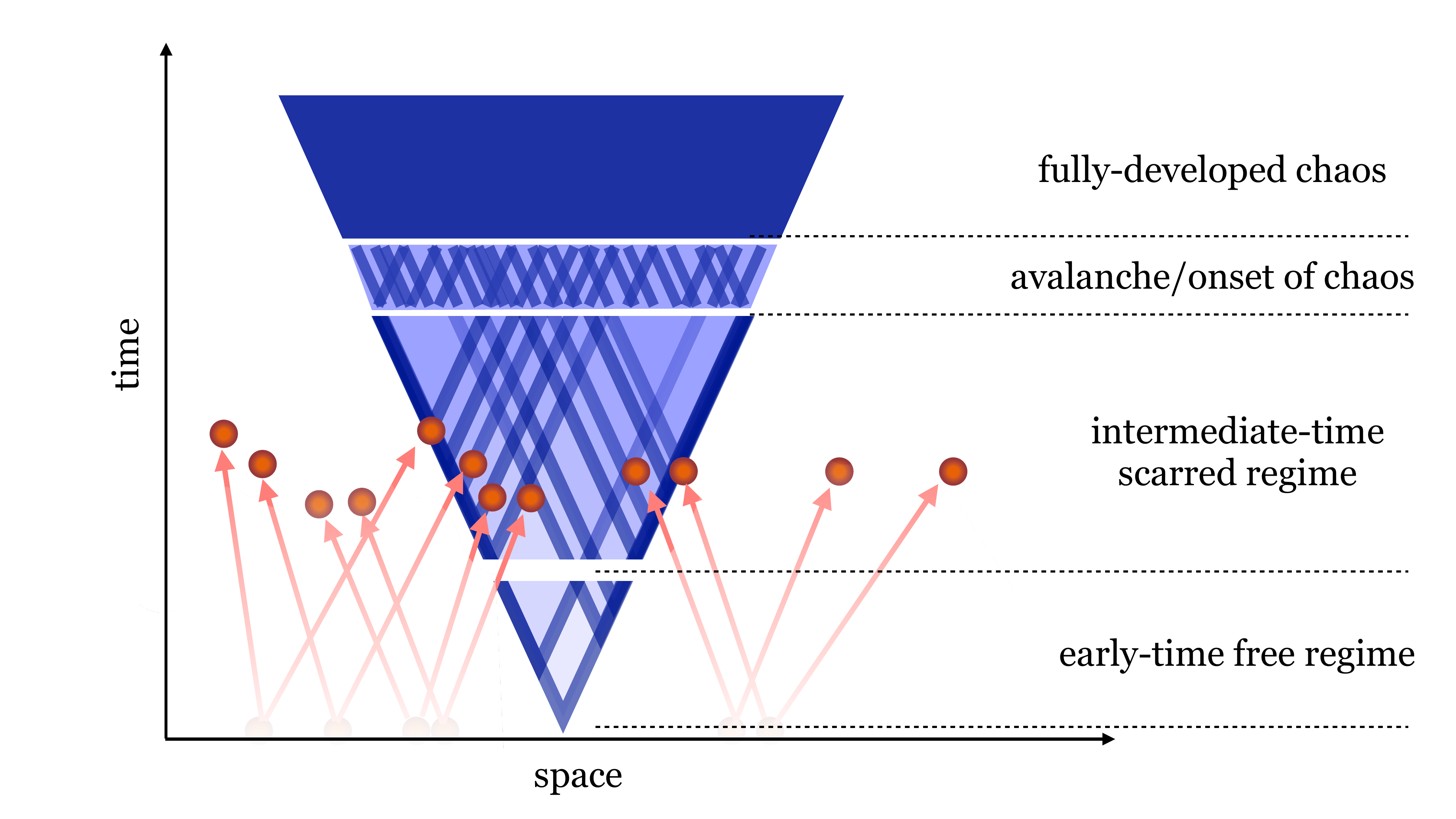}
\caption{
Schematic diagram of the different regimes leading to the onset of many-body chaos, as measured by the decorrelator (Eq.~\ref{eq:DecorrMain}), in classical unfrustrated spin systems. The initial state can be viewed as a (quasi) long-ranged ordered state interspersed by disordered `defect' regions; this constitutes a proxy for a low-temperature initial state where the disordered regions can be thought of as mimicking the excitations on top of a ground state. The quasiparticles corresponding to the excitations, denoted by the red circles, scatter off the primary lightcone of spatiotemporal chaos, seeding secondary lightcones inside the primary one. At intermediate times, this leads to a scarred regime, following which the secondary lightcones proliferate and further scatter off each other, leading to fully-developed chaos at late times.}
\label{fig:schematic}
\end{figure}

Indeed, the spatiotemporal structure of many-body chaos~\cite{livi1986distribution,cencini2001linear,vastano1988information,lepri1996chronotopic,lepri1997chronotopic,giacomelli2000convective} in classical and quantum many-body systems has received a lot of attention recently~\cite{aleiner2016microscopic,shenker2014black,maldacena2016bound,bohrdt2017scrambling,luitz2017information,rozenbaum2017lyapunov,hosur2016chaos,kurchan2016quantum,roberts2016lieb,das2018light,bilitewski2018temperature,bilitewski2021classical,ruidas2021many,PhysRevLett.127.124501,banerjee2025intermittent,PhysRevE.89.012923,PhysRevLett.112.210601,iyoda2017scrambling,swingle2017slow,banerjee2017solvable,scaffidi2017semiclassical,PhysRevX.7.031047,patel2017quantum,PhysRevX.8.031057,PhysRevX.8.031058,von2017operator,nahum2017operator,PhysRevB.97.144304,PhysRevLett.119.026802,Schuckert_2019,de_Wijn_2013,PhysRevLett.109.034101,AltmanScrambling2020,scaffidi2017semiclassical,PhysRevX.9.031048,PhysRevLett.129.160601,PhysRevLett.132.030402}.  These studies have focused on the characterization of the temporal and spatial aspects via the Lyapunov exponent, $\lambda_L$, and butterfly velocity, $v_B$ respectively. The {\it out-of-time commutators} (OTOCs)~\cite{1969LarkinOvchin,kitaev} and their classical analogue, the {\it decorrelators}~\cite{das2018light} reflect the time and length-scales of such spatiotemporal chaos. The classical decorrelators, $\mathcal{D}(i,t)$ (Eq. \ref{eq:DecorrMain}), measure the evolution of the difference of two configurations which initially ($t=0$) are infinitesimally different only at the origin ($i=0$).
In the chaotic regime, this difference grows with a velocity-dependent Lyapunov exponent which is positive/negative inside/outside a lightcone, respectively -- this is a spatiotemporal representation of the celebrated butterfly effect~\cite{lorenz2000butterfly,lorenz1996essence,hilborn2004sea}. The inside of the lightcone defines the butterfly velocity $|i|<v_B t$; here,  as indicated in Fig. \ref{fig:schematic}, the decorrelator grows exponentially, exhibiting the extreme sensitivity to initial conditions~\cite{gaspard2005chaos,lichtenberg2013regular,schuster1988deterministic,kanekobook}.

The fully developed chaotic regime is typically preceded by an early time non-exponential, but ballistically spreading, decorrelator as schematically shown in Fig. \ref{fig:schematic}. The nature of this initial regime, alongside its crossover to fully-developed chaos, therefore, carries key information on the phenomenon of the {\it onset} of many-body chaos~\cite{van1982album,Chowdhury_2017}. 

Unlike that of typical difference maps of chaos such as the logistic map~\cite{ausloos2006logistic} where the onset only occurs beyond a certain regime of the parameters, in interacting many-body systems, it is the non-linear terms in the equation of motion that generically lead to chaos via coupling of harmonic normal modes somewhat similar to chaotic scatterings~\cite{gaspard2005chaos} of a particle, but in a dynamically evolving environment. However, a suitable choice of an ensemble from which the initial conditions are drawn -- one where non-linear effects are initially {small} -- allows for a controlled study of the onset of chaos. In this situation, two smooth initial conditions decorrelate slowly at early times, only gradually giving way to a faster and ultimately exponential form of the decorrelator as the effect of the non-linear scatterings accumulates. 

A natural setting for such a separation of timescales is a dilute gas of weakly interacting (quasi-)particles with long mean-free paths. In a condensed matter setting, this generically occurs as a consequence of the spontaneous breaking of a (continuous) symmetry, resulting in long-lived Goldstone modes~\cite{chaikin2000principles}; but also for other reasons, as is famously the case for Landau quasiparticles in a Fermi liquid \cite{nozieres1969theory}. Such almost harmonic low temperature settings provide platforms for the observation of the various stages of spatiotemporal chaos.

In this paper, we consider classical Heisenberg spin chains as model systems, as their spin-wave-like excitations have decreasing ($\sim \sqrt{T}$) amplitudes at low temperature, naturally suppressing non-linear effects in a tunable fashion.
Building on the results of Ref. \cite{bilitewski2021classical}, we examine three qualitatively and quantitatively distinct regimes in the decorrelator, schematically drawn in Fig. \ref{fig:schematic} preceding the final regime of fully-developed chaos. At the earliest times, the effect of the deviations from the perfectly ordered spin configuration can be neglected locally, such that the evolution of the {\it free} decorrelator is captured by a spread of a localised wave-packet in a frozen magnetically ordered background. Truncating the equation of motion for the decorrelator at harmonic orders gives a quantitatively accurate description of this regime. As the decorrelator evolves and spreads, it senses the effect of deviation from the perfect order. This is captured by the mode-coupling between the decorrelator and the harmonic spin-wave dynamics of the local deviations. This seeds sharp streaks --  {\it secondary lightcones} -- at the boundary of the  (primary) decorrelator lightcone, leading to the second stage, where the inside of the lightcone appears {\it scarred} (see Fig. \ref{fig:lc_scarred} for numerical data) and heterogeneous. The increasing number of secondary lightcones and their mutual scattering lead to the cascade of tertiary and higher generation lightcones inside the primary one. Notably, the tertiary and higher generation lightcones are triggered by the scattering between spin waves due to the underlying non-linear anharmonic terms, and hence their appearance signals the breakdown of the harmonic spin-wave approximation.  Indeed, the magnitude of the decorrelator grows rapidly around isolated {\it hotspots} within the primary lightcone. The number of such hotspots grows exponentially in the third and final stage of the onset of chaos, ultimately leading to the fully developed chaotic regime.

To reveal the different stages of the onset for the spin chain, we choose initial states from a {\it defect ensemble} defined as follows (see Sec. \ref{sec:defect-ensemble} and Appendix \ref{appen:ThermalCorrespondence} for details). Starting with the fully ordered state (fully polarised ferromagnet or antiferromagnetic Neel state), we generate an ensemble of configurations where on a fraction, $\rho_d$, of sites,  the spins are randomly oriented away from the direction of ordering. The strictly on-site nature of the defects restricts the spin precessional dynamics (Eq. \ref{eq:HeisenbergModel}), at early times, around the defect sites in an otherwise frozen background and allows a quantitative investigation of the origin of the secondary lightcones and different stages of the onset. In particular, explicit solutions of the mode-coupling equations derived in Ref. \cite{bilitewski2021classical} for a single defect (Fig. \ref{fig:single-defect-FM}) trace the origin of the secondary lightcone to the dynamics of the defects, which generate their individual {\it spin-dynamics lightcones}. With time, these spin-dynamics lightcones then spread and reach the edge of the primary decorrelator lightcone at well-defined points in space-time, seeding the secondary lightcone inside the primary one. 

{We quantify the heterogeneity caused by the secondary lightcones and their scattering within the primary one via an appropriately defined (inverse-) participation ratio for the normalised decorrelator, $g(t)$ (Eq. \ref{eq:gt_main}).} The time dependence of $g(t)\sim t^{a}$ reflects the different regimes (Fig. \ref{fig:gt_med_compare}) via the exponent $a (\geq0)$. 
The temporal extents of the different regimes, as evidenced by $g(t)$, depend on the initial defect density, $\rho_d$.

The early time free evolution of the decorrelator corresponds to defect-density-independent regime (Fig. \ref{fig_tfree}), but its duration scales with defect density as $\rho_d^{-1/2}$. {As more secondary lightcones are formed, scattering amongst them increases in importance, accelerating the growth of the decorrelator around the scattering space-time points. In the ensuing second regime, the decorrelator is dominated by a few such points at which it has {parametrically} large magnitude in a background which otherwise mirrors the decorrelator for a defect-free, ordered initial state. This results in a plateau-like feature in $g(t)$, i.e.\ a small value of $a$, Fig. \ref{fig:gt_med_compare}.  

With time, however, the increasing density of the secondary lightcones leads to their overlap, rapidly amplifying the effect of the non-linearity. 
In particular, these non-linear effects almost coincide with the appearance of the tertiary and a cascade of higher generation of lightcones. This leads to exponential amplification of the decorrelator around the hotspots, which provides positive feedback to the growth of the decorrelator locally around these sites. This ultimately leads to chaos. The mechanism is reflected in the fat-tailed distribution, $\mathcal{P}(\mathcal{D})$ of the decorrelator (Fig. \ref{fig:DDist}). The exponentially growing number of hotspots truncates the plateau in $g(t)$ (Fig. \ref{fig:gt_med_compare}) via a sharp knee (Eq. \ref{eq:gt-1/At} and Fig. \ref{fig:AtKtgt}). 

The fully developed chaotic regime is characterised by $\lambda_L$ and $v_B$, which can be extracted from the decorrelator (Eq. \ref{eq:Dnorm-postulate}) via the velocity-dependent Lyapunov exponent (VDLE) (Eqs.~\ref{eq:vdle-def} and \ref{eq_vlde}). While the Lyapunov exponent $\lambda_L$ increases with the increasing defect density $\rho_d$, the butterfly velocity $v_B$ decreases. This compares favourably with the characterisation of chaos with increasing energy density in thermal ensembles. At even longer times, the full decorrelator saturates, leading to a homogeneous lightcone, such that at long times $g(t)\sim t^{-1}$. The saturation can be indefinitely delayed by a decrease in the size of the initial perturbation; in the limit of vanishing perturbation, studying the {\it `linearised'} decorrelator (Eq. \ref{eq:lin-decorr}) {eliminates the late-time saturation}. While the decorrelator still spreads ballistically with $v_B$, almost its entire weight (the core) spreads in a sub-ballistic fashion (Fig. \ref{fig:contours}).

 The emergence of the inner core is rooted in the statistics of the timescales required for a spatial point to transition into the fully developed exponential-chaos regime. Consider a point $i$ reached by the primary lightcone at time $t = i/v_B$, where the decorrelator remains small, $\mathcal{D}(i, i/v_B) \ll 1$. The subsequent crossover to exponential growth does not occur immediately but proceeds through a sequence of onset stages whose timing is dictated by the statistics of defect dynamics. Scattering of defects off the primary lightcone generates secondary and higher-order lightcones, each further amplifying the decorrelator. Points closer to the origin encounter statistically more such higher-order lightcones than points farther away, and therefore enter the exponential regime earlier. This spatial variation in the crossover times produces the inner core, while the underlying statistics of scattering events, as discussed in Sec.~\ref{subsec_core}, determine its sub-ballistic expansion.

The rest of the paper is organised as follows. We start by concretely describing the particular setting of our calculations, including the Hamiltonians, the defect ensemble, the summary of the decorrelator and essential numerical details in Sec. \ref{sec_prelim}. This is followed by a detailed quantitative characterisation of the four different states of the decorrelator, from free evolution to fully-developed chaos, as well as the origin of the secondary lightcones in Sec. \ref{sec_anatomy}. Finally, in Sec. \ref{sec_lypvb} we study the form of the decorrelator in the fully developed chaotic regime and show the dependence of both the Lyapunov exponent and the butterfly velocity on the defect density. We conclude with a summary of our results in Sec. \ref{sec_summary}. Various technical details are discussed in the appendices. 
 
\section{Preliminaries and definitions}
\label{sec_prelim}

We start with laying out the basic definitions and describing the central setting used throughout this work. 
Specifically, in the following, we describe the model, the ensemble of initial states, and the formalism of the classical decorrelator as a measure of spatiotemporal chaos.

\subsection{Classical Heisenberg spin chain}
To put our ideas on a concrete footing, we will employ the well-studied
classical Heisenberg spin chains consisting of three-component classical spins, ${\bf S}_i=(S^x_i, S^y_i, S^z_i)$ (with $|{\bf S}_i|=1$) at each lattice site and interacting via nearest-neighbour spin-spin  interactions, given by the Hamiltonian
\begin{equation}
    \mc{H} = -J\sum_{\langle ij \rangle} \mb{S}_i \cdot \mb{S}_{j}
    \label{eq:HeisenbergModel}
\end{equation} 
where $J$ is the strength of the ferromagnetic $(J > 0)$ or anti-ferromagnetic $(J < 0)$ interaction. For the former, the lowest energy configuration consists of spins oriented in parallel along any particular direction on the Bloch sphere, while for the latter neighbouring spins are antiparallel, forming the Neel state. The classical precessional dynamics is given by the Landau-Lifshitz-Gilbert equation 
\begin{equation}
    \frac{\d\mb{S}_j}{\d t} = \mb{S}_j \times J\big(\mb{S}_{j-1} + \mb{S}_{j+1}\big). \label{eq:spin_dynamics}
\end{equation} 
The dynamics conserves the total spin, ${\bf S}_T=\sum_{i}{\bf S}_i$, in addition to energy.

\subsection{Classical decorrelator}
To characterise the spatiotemporal chaos quantitatively, we utilise the recently developed formalism of the decorrelation function \cite{das2018light, bilitewski2018temperature, bilitewski2021classical, ruidas2021many} defined as \begin{equation}
    {\cal D}(i, t) = \big(\mb{S}^a_i(t) - \mb{S}^b_{i}(t)\big)^2, \label{eq:DecorrMain}
\end{equation} 
where $\mb{S}_i^{a(b)}$ are the spins on the two copies, copy-$a$ and copy-$b$ of the system.
The two copies are initialised in an identical state except at a single seed site where they differ infinitesimally,
such that, 
\begin{equation}\mb{S}_i^{b} = \mb{S}_i^{a} + \delta \mb{S}_0\, \delta_{i, 0}\,;\quad\delta \mb{S}_0 = \varepsilon (\hat{\mb{n}} \times \mb{S}_0^a)\,,\label{eq:perturb}
\end{equation} 
where $\hat{\mb{n}} = (\hat{e}_{x}\times \mb{S}^a_0)/|\hat{e}_{x}\times \mb{S}^a_0|$. Physically, this means that the spin at the site $i=0$ of copy-$b$ is infinitesimally (of strength $\varepsilon$) rotated about the unit vector $\hat{\mb{n}}$ relative to that of copy-$a$. The choice of $\hat{\mb{n}}$ is general as long as we get a non-zero perturbation at the seed site. As a matter of notation, for a system with $N$ spins, we label the sites with integers $(-N/2, N/2]$ and the seed site is chosen to be the origin $i=0$. We will study the decorrelator dynamics averaged over an ensemble of initial states. The ensemble is described in Sec.~\ref{sec:defect-ensemble} and we denote the averaged decorrelator via $\braket{{\cal D}(i,t)}$.

Note that the decorrelator, as defined in Eq.~\ref{eq:DecorrMain}, is bounded between $0$ and $4$. At the same time, at $t=0$, the decorrelator at the seed site ${\cal D}(i=0,t=0)\sim \varepsilon^2$.
Therefore, for any finite $\varepsilon$ however small (as required in the numerical calculations), there exists a temporal window only parametrically large in $\varepsilon^{-1}$ over which the decorrelator can show an exponential growth before it saturates. 
To extract a Lyapunov exponent in a meaningful way, we therefore also study the decorrelator directly in the limit of $\varepsilon \to 0$.
Defining $\mb{z}_i(t) = \lim_{\varepsilon\rightarrow0}\delta \mb{S}_i(t)$, a linearised equation in this limit can be written as 
\begin{equation}
    \dot{\mb{z}}_i = J \mb{S}_i \times (\mb{z}_{i-1} + \mb{z}_{i+1}) + J
    \mb{z}_{i} \times (\mb{S}_{i-1} + \mb{S}_{i+1})\,, \label{eq:LinearDyn}
\end{equation}
where the spin dynamics for $\mb{S}_j(t)$ is calculated via the full non-linear evolution in Eq.~\ref{eq:spin_dynamics}.
Naturally, the solution for $\mb{z}_i(t)$ does not exhibit a saturation and grows unboundedly with time.
The ensemble-averaged linearised decorrelator is given simply by 
\begin{equation}
	\braket{{\widetilde{\cal D}}(i,t)} = \braket{|\mb{z}_i(t)|^2}\,,
	\label{eq:lin-decorr}
\end{equation}
which grows exponentially at late times. 
This allows us to define the velocity-dependent Lyapunov exponent (VDLE)~\cite{khemani2018velocity} as 
\begin{equation}
	\lambda_L(v)\equiv \lim_{t\rightarrow\infty} \ln(\braket{\widetilde{\cal{D}}(i=vt,t)})/2t\,.
    \label{eq:vdle-def}
\end{equation}
For classical systems of the kind considered here, the VDLE is expected to be positive inside the lightcone $(v<v_B)$ and negative outside it $(v>v_B)$, indicating the exponential growth and decay with time of the decorrelator, respectively, in the two cases. In fact, the velocity where $\lambda_L(v)$ changes sign is a useful estimate of $v_B$.

\subsection{The `defect' ensemble of initial states  \label{sec:defect-ensemble}}
Our ensemble of initial states, which we refer to as the \textit{defect ensemble}, is defined as follows. 
Starting from a perfectly aligned ferromagnetic or antiferromagnetic configuration, we randomly select $N_d$ \textit{defect sites}, corresponding to a defect density $\rho_d \equiv N_d / N$. 
These sites, labelled by indices $\{ i_d \}$, host spins that are rotated relative to the order direction (local in case of antiferromagnet) by a polar angle $\theta_{i_d}$ and an azimuthal angle $\phi_{i_d}$, where each $\theta_{i_d}$ is drawn independently from a distribution such that $\cos\theta_{i_d}$ is sampled uniformly in $[1,\cos\theta_M]$, and $\phi_{i_d}$ is sampled uniformly in $[0, \phi_M]$. 
The ensemble of states generated by varying both the choice of defect sites and the rotation angles $\{ \theta_{i_d}, \phi_{i_d} \}$ thus defines our \textit{defect ensemble}, characterised by three parameters $(\rho_d, \theta_M, \phi_M)$.

At this stage, it is important to discuss the motivation behind this construction of the defect ensemble of initial configurations. In thermal equilibrium at low temperature, typical spin configurations deviate from the perfectly ordered ferro- or antiferromagnetic ground state due to a finite density of spin-wave excitations spanning a range of wavelengths. 
These excitations correspond to transverse fluctuations of the spins about the direction of perfect alignment. 
Previous work~\cite{bilitewski2021classical} 
demonstrated that interactions between the decorrelator and thermally excited spin-waves give rise to \textit{secondary lightcones} -- the hallmark of the scarred regime that precedes the onset of chaos. 
However, because thermal fluctuations involve a continuum of coupled spin-wave modes, it is difficult to quantitatively trace the emergence of these secondary lightcones or to isolate the microscopic mechanism by which spin-wave interactions generate chaotic dynamics.

Our construction of the defect ensemble circumvents this difficulty. We replace the thermal ensemble with an ensemble of states containing a finite density of randomly oriented single-spin defects 
such that at early times the dynamics of the system is localised to the neighbourhood of the defect spins. Subsequent overlap of these dynamical regions (spin-dynamics lightcones) activates the underlying non-linear scattering of the spin-waves and thus separates different aspects of the physics of the crossover to chaos. This simplification preserves the essential ingredients, namely, local transverse deviations from perfect order, while allowing us to systematically track how the resulting perturbations evolve in space and time and eventually interact with the primary lightcone of the decorrelator. 
In this way, the defect ensemble provides a minimal, tractable, and tunable setting in which to uncover the microscopic origin of secondary lightcones in the decorrelator and to elucidate the transient scarred regime that ultimately leads to fully developed spatiotemporal chaos. 
We expect the underlying mechanism to be the same as in the thermal case, with the defects effectively mimicking the role of finite temperature. In particular, for $\rho_d$ approaching $1$ and small $(\theta_M, \phi_M)$, the defect magnitude varies almost continuously -- akin to thermal fluctuations, albeit at short wavelengths. 

In Appendix~\ref{appen:ThermalCorrespondence}, we characterise these configurations {\it vis-{\`a}-vis} the thermal ensemble.
However, for much of the numerical results in the following, we set $\theta_M=\pi$ and $\phi_M=2\pi$ unless otherwise stated, and study the physics as a function of $\rho_d$.

\subsection{Methodology}

Before delving into the results, we briefly outline the general methodology of our analysis.
For initial conditions drawn from the `defect' ensemble discussed above, we simulate the Hamiltonian precessional dynamics using a fourth-order Runge-Kutta (RK4) method, where we set $J=1\,(-1)$ for the ferromagnet (antiferromagnet). The numerical time step $\Delta t = 0.005$ is chosen such that the usual $O(\Delta t^4)$ errors conserve total energy and spin to our acceptable precision.
The two copies, $a$ and $b$, with the initial condition being different at a single site (see Eq.~\ref{eq:perturb}) are evolved independently using the RK4 dynamics and the decorrelator is calculated from the difference in Eq.~\ref{eq:DecorrMain}.
For the numerical computation of the linearised decorrelator, the spin dynamics in a single copy is solved for using the RK4 method {\it without} any perturbation. The solution for $\{\mb{S}_i(t)\}$ is then fed back into Eq.~\ref{eq:LinearDyn} to solve for the dynamics of $\mb{z}_i(t)$.
This numerical analysis constitutes the core method for obtaining many of the results in this work. 

However, as motivated earlier, a key ingredient in our theory for the onset of chaos is the scattering of the dynamical defects in the initial conditions off the primary decorrelator lightcone. 
To get an analytical handle on how they seed secondary decorrelator lightcones, we use a mode-coupling theory~\cite{bilitewski2021classical} adapted to the present setting as follows. 
In a given initial spin-configuration, each spin can be decomposed as 
\begin{align}
    \mb{S}_i(t) = \zcap\sqrt{1 - |\mb{L}_i(t)|^2} + \mb{L}_i(t)\,, \label{eq:SW_Expansion}
\end{align}
where $\mb{L}_i(t)$ is the component perpendicular to the order direction ($ \mb{L}_i \perp {\hat{e}_{z}}$), taken to be ${\hat{e}_{z}}$ here without loss of generality. 
Simplifying to the case of a single defect such that 
\begin{equation}
	\mb{L}_i(t=0) = |\mb{L}_d|\delta_{i,i_d}\,,
\end{equation}
where $\mb{L}_d = (\sin\theta_d\cos\phi_d, \sin\theta_d\sin\phi_d, 0)$ and focusing on the FM case, the defect dynamics can be solved for up to linear order, leading to 
\begin{multline}
    {\bf L}_i(t) = |\mb{L}_d|\, \mathcal{J}_{|i - i_d|}(2 t) \big[\cos(\phi_d - 2 t + \pi |i - i_d|/2) \hat{e}_{x} \\ + \sin(\phi_d - 2 t + \pi |i - i_d|/2)\hat{e}_{y}\big]\,,
    \label{eq:defect-dynamics-linear}
\end{multline}
with $\mathcal{J}_i(t)$ denoting the Bessel function of the first kind.
The above expression encodes the spreading of the defect in space with time, and acts as a proxy for the dynamics of {thermally excited spin-waves} in a finite-temperature state.

Defining $\dS_i(t) \equiv \mb{S}_{i}^{b}(t) - \mb{S}_{i}^{a}(t)$, the dynamics of the decorrelator in Eq.~\ref{eq:DecorrMain} can now be expressed through the equation for the precessional dynamics (Eq.~\ref{eq:spin_dynamics}) as 
\begin{multline}
    \frac{\d\,  \dS_i(t)}{\d t} = \zcap \times \bigg( \sum_{j \in i} J \dS_j - 2 J\dS_i\bigg) \\+ \mb{L}_i \times \sum_{j \in i} J \dS_j + \dS_i \times \sum_{j\in i} J \mb{L}_j \\ - \frac{1}{2}\zcap\times\bigg(|\mb{L}_i|^2 \sum_{j \in i}J\dS_j - \dS_i\sum_{j \in i}J|\mb{L}_j|^2\bigg), \label{eq:diffNLin_main}
\end{multline}
up to the second order in spin-wave amplitude $|\mb{L}_i|$.
Analysing the above equation with the form of the defect dynamics from Eq.~\ref{eq:defect-dynamics-linear} as input to it, as we elaborate later, sheds insights into the origin of the secondary lightcone.


\section{Anatomy of the lightcone: crossover to chaos}
\label{sec_anatomy}

We now turn to the main focus of this work, namely, to understand how the collective effect of several defects dynamically scattering amongst themselves, eventually leads to fully-developed chaos with an intermediate scarred regime en route. To address this, here, we present a detailed analysis of the anatomy of the decorrelator in the presence of finite defect density and identify different temporal regimes. As mentioned in Sec.~\ref{sec:defect-ensemble}, we will consider initial states drawn from the defect ensemble characterised by $(\rho_d,\theta_M,\phi_M)$. Unless stated otherwise, the results in this section are for $\theta_M=\pi$ and $\phi_M=2\pi$.

 To first present a broad brush view of the spatiotemporal structures inside the lightcone, in Fig.~\ref{fig:lc_scarred}, we show the decorrelator as a density plot for a {\it single} initial configuration drawn from the defect ensemble. 
Three qualitatively distinct temporal regimes are evident in the plots for both the FM and the AFM case. At early times, the decorrelator is devoid of any effect of the defect sites and the spin-dynamics lightcones emanating from them; this is the regime where, in a statistical sense, none of the spin-dynamics lightcones -- emanating from the defect sites -- have yet reached the edges of the primary lightcone. This results in the free evolution of the decorrelator similar to the ultrashort-time integrable region in Ref.~\cite{bilitewski2021classical}. 

This free decorrelator evolution is followed by the scarred regime where the secondary lightcones seeded by the defects hit the edge of the spreading primary decorrelator lightcone, resulting in visible streaks of the former inside the latter. With time, the number of such secondary lightcones and their intensity increase such that they are not only visible but also are the dominant feature inside the lightcone, giving rise to tertiary lightcones within the primary one due to the scattering of the secondary lightcones. {Notably, the scattering arises from the non-linear terms in the equation of motion (Eq. \ref{eq:spin_dynamics}). In this sense, the appearance of the tertiary lightcones due to the overlap of the secondary ones activates the non-linear terms, setting up an {avalanche} process.  Finally, at even later times, these secondary, tertiary and higher-order lightcones grow further and proliferate to give way to the fully-developed chaos indicated by a statistically uniform decorrelator at its saturation value inside the lightcone. The heatmaps in Fig.~\ref{fig:lc_scarred} can be considered the quantitative counterpart to the schematic shown in Fig.~\ref{fig:schematic}, obtained for the classical Heisenberg spin chain.

\begin{figure}
	\centering
	\includegraphics[width=1.0\linewidth]{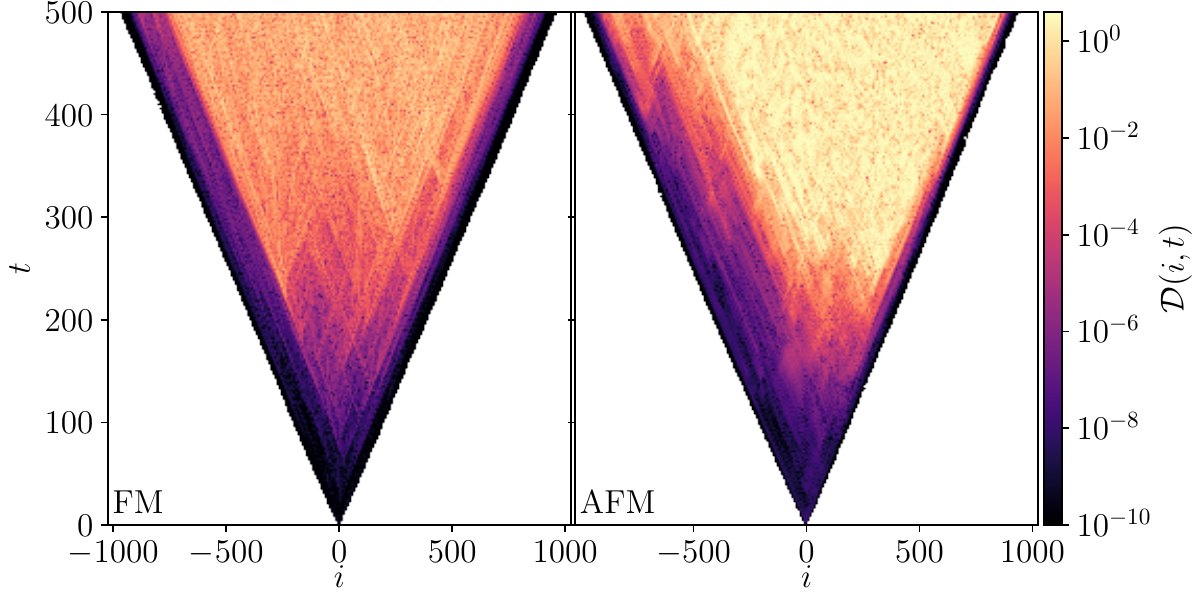}
		\caption{The decorrelator, ${\cal D}(i,t)$, shown as a heatmap (with logarithmic colour-scales) in space-time for a single initial configuration drawn from the defect ensemble. The scarred regime, indicated by the streaky patterns inside the lightcone, at intermediate times arising out of the secondary lightcones, is clearly visible along with their proliferation at late times, leading to chaos. The left and right panels correspond to the FM and AFM cases, respectively. For these plots $\rho_d=0.05$, $\theta_M=\pi$, $\phi_M=2\pi$ and $\varepsilon=10^{-4}$ with $L=2048$.} 
	\label{fig:lc_scarred}
\end{figure}

\begin{figure}
	\centering
	\includegraphics[width=0.9\linewidth]{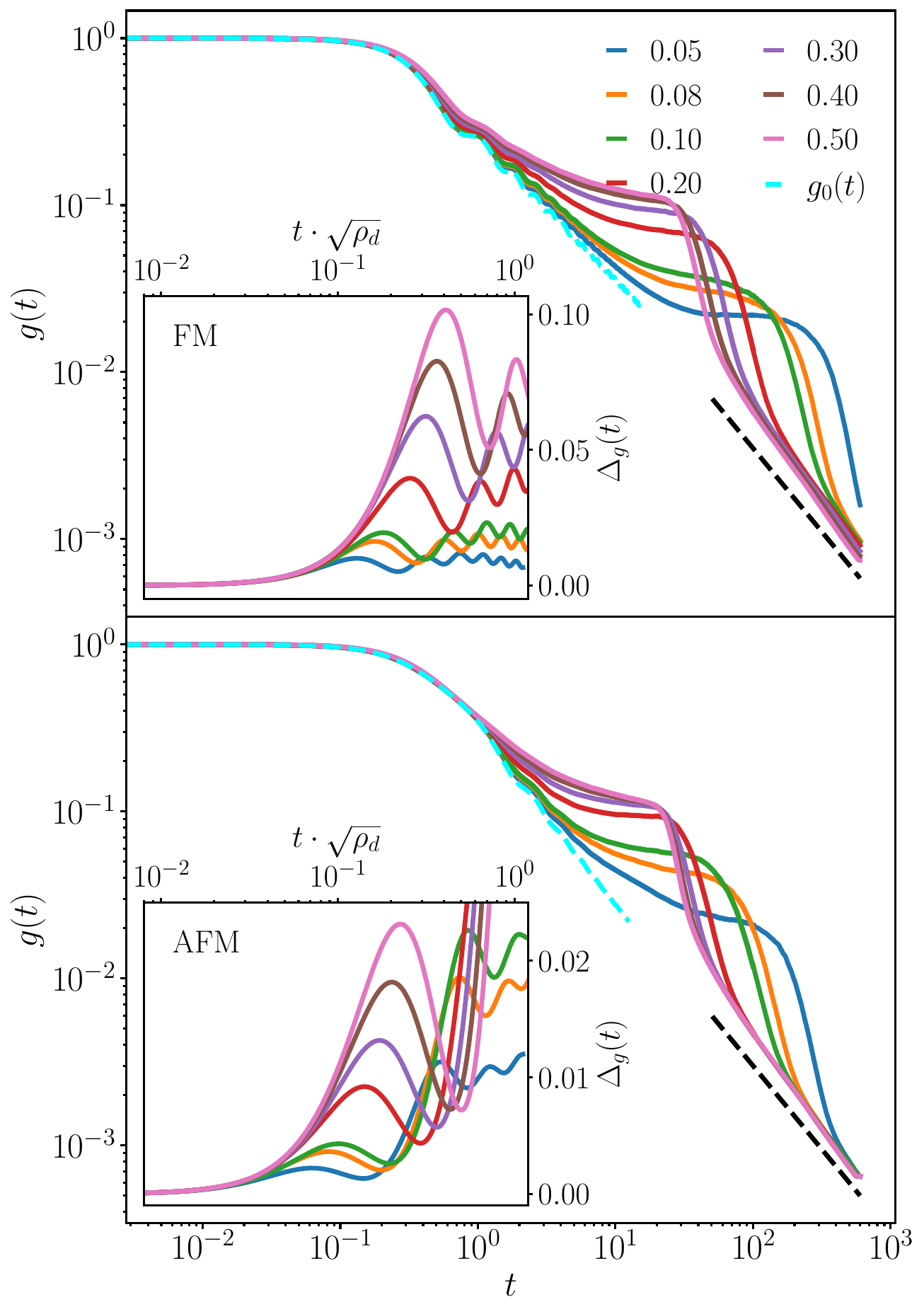}
		\caption{The participation function $g(t)$ (defined in Eq.~\ref{eq:gt_main}) plotted for different defect densities $\rho_d = 0.05,\ldots,0.50$ for ferromagnetic (upper panel) and antiferromagnetic (lower panel) couplings. The defect-free behaviour for the respective cases is shown in a cyan dashed line. For each defect density, we can distinguish three different regimes. Within the early times (up to ${O}(1)$), $g(t)$ coincides with the defect-free behaviour $g_0(t)$ (dashed cyan line). The onset of chaos in the intermediate time regime, as denoted by the plateau and the avalanche decay in $g(t)$, happens at the same time for FM and AFM cases, and therefore is independent of the sign of the interaction between the spins. Thereafter, the chaotic regime is indicated by $g(t) \sim 1/t$ as denoted by the black dashed line. In the insets, we show the scaling of the deviation $\Delta_g(t) = |g(t) - g_0(t)|$ from the free behaviour with defect density $\rho_d$. For these plots, $\theta_M = \pi$, $\phi_M = 2\pi$, system size $N = 2048$ and the results are averaged over 10240 configurations.} 
	\label{fig:gt_med_compare}
\end{figure}

The above discussion makes it evident that the key to understanding these different regimes quantitatively is via the statistical properties of the decorrelator over space in these different temporal regimes. In particular, to characterise the possible heterogeneity of the decorrelator due to the secondary lightcones quantitative, we define 
\begin{equation}
    g(t) = \left\langle\frac{\sum_j {\cal D}^2(j, t)}{\Big(\sum_j {\cal D}(j, t)\Big)^2}\right\rangle\,, \label{eq:gt_main}
\end{equation}
 which can be understood as an inverse participation ratio of the `normalised' decorrelator over space. The function $g(t)$ has one physical interpretation as follows.
At any given time $t$, the decorrelator will be supported over a region of length $\propto t$ due to the finite butterfly velocity.
Therefore, if the decorrelator is statistically homogeneous inside the lightcone, $g(t)\sim t^{-1}$. 
On the other hand, if the decorrelator values are dominated by $O(1)$ number of relatively large values inside the lightcone, $g(t)\sim t^0$. 
An intermediate scaling of $g(t)\sim t^{-\alpha}$ with ${0<\alpha<1}$ indicates a non-trivial distribution of the ${\cal D}(i,t)$ over space.

In Fig.~\ref{fig:gt_med_compare} we present the results for $g(t)$ as a function of time $t$ for both the FM and AFM, for different defect densities $\rho_d$. Again, distinct temporal regimes are clearly visible. In the following, we discuss these regimes in detail one by one in accordance with the schematic Fig. \ref{fig:schematic}.

\subsection{Early-time free regime}
 
The early time regime is indicated by the absence of any dependence of $g(t)$ on the defect density $\rho_d$. 
In fact, in this regime, the function $g(t)$ follows the no-defect case, denoted by $g_0(t)$ where the latter is given by Eq.~\eqref{eq:gt_main} with ${\cal D}(i,t)$ given by the Bessel function solution for the FM and an integral form for the AFM case, as detailed in Ref.~\cite{bilitewski2021classical}.
This is denoted by the cyan dashed lines in Fig.~\ref{fig:gt_med_compare}.
As mentioned earlier, this is the regime where the times are short enough such that the spin-dynamics lightcones of none of the defects reach the edges of the primary decorrelator lightcone. The decorrelator in this regime is then identical to that of the case of a fully ordered initial state. 
This implies that the free temporal regime extends up to shorter timescales for larger defect densities. 
That this is indeed the case is evident in the data in Fig.~\ref{fig:gt_med_compare} as the curves of $g(t)$ for higher $\rho_d$ deviate from those of $g_0(t)$ at earlier times. 
A local and arguably more direct probe of the above physics is the comparison of the decorrelator at the origin for a finite defect density with that of the no-defect case. This is shown in Fig.~\ref{fig_tfree}(left), where the cyan dashed line denotes the decorrelator at the origin in the absence of any defect. The decorrelators for finite defect densities (solid lines) follow the no-defect case at early times and deviate from it at a timescale which decreases with increasing defect density.

The free evolution for the decorrelator continues up to a time, $t_{\rm free}(\rho_d)$, till the first defect hits it. The distribution of this time is governed by the conditional probability that the nearest defect occurs at $i_0$ from the origin (either left or right) and is given by
\begin{align}
    P(i_0)=2\frac{^{N-2i_0}C_{N_d-1}}{^NC_{N_d}}\,,
    \label{eq_probtfree}
\end{align}
where $N_d=\rho_d L$ and the factor $2$ appears because the nearest defect can occur on either side of the origin. From the above distribution, the distribution of $t_{\rm free}(\rho_d)$ can be obtained by using $t_{\rm free}(\rho_d)=i_0/(2v_B)$ and setting $v_B=2$ at these early times of free propagation of the decorrelator and the defect magnetisation. The resultant comparison is shown in Fig. \ref{fig_tfree} (right) for $\rho_d=0.01, 0.05$ and $0.1$ for $N=1024$ with $\approx 10^4$ configurations.

The dependence of $t_{\rm free}(\rho_d)$ on the defect density is further evident from the collapse of the difference $\Delta_g(t)\equiv |g(t)-g_0(t)|$ under the rescaling $t\rightarrow t\sqrt{\rho_d}$ as shown in the insets of Fig.~\ref{fig:gt_med_compare}. The particular scaling can be understood as a geometric effect of the area within the decorrelator lightcone that is getting filled up by the secondary ones, causing its departure from the free value. The deviation of the decorrelator, $\Delta {\cal D}$, at a given time-slice occurs over a spatial region within the primary lightcone that scales as $\sim t$ while the number of defects that contribute to $\Delta {\cal D}\sim t\rho_d$ (where we have neglected the contributions from the tertiary or higher-order lightcones and dependence of $v_B$ on $\rho_d$ at these early times) such that
\begin{align}
    \Delta g \sim \Delta {\cal D}\times t\sim t^2\rho_d\,,
    \label{eq_deltag}
\end{align}
which explains the $\sqrt{\rho_d}$ scaling of the timescale of free evolution.

\begin{figure}
	\includegraphics[width=1.0\linewidth]{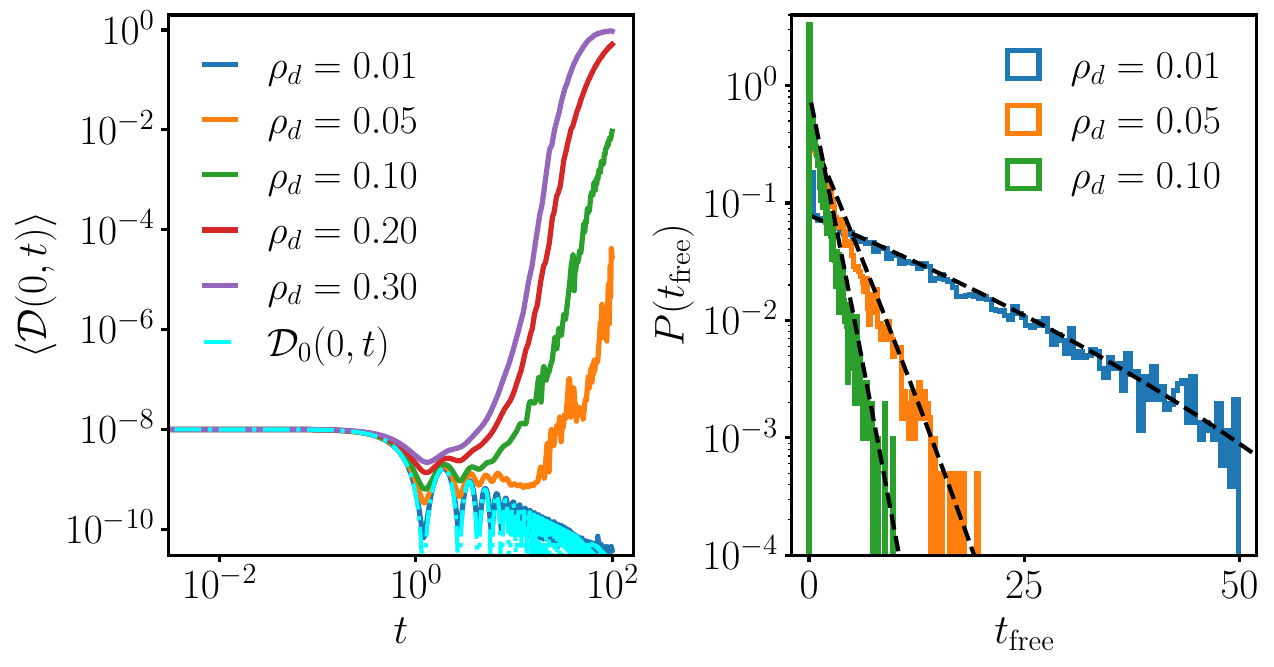}
		\caption{Early-time free behaviour of the decorrelator for FM. Left: On-site time evolution of the decorrelation function at the $i=0$ site for different defect densities. The cyan dashed line denotes the defect-free case given by the Bessel function solution. The finite defect curves follow the free behaviour up to the time $t_\mr{free}$. Right: The distribution of time of free propagation for the decorrelator obtained numerically for $N=1024$ with $10240$ configurations. The black dashed lines denote the analytical results obtained from $t_{\rm free}=i_0/(2 v_B)$ and using the distribution of $i_0$ in Eq. \eqref{eq_probtfree}. Given that the lowest probability we can attain is $\mc{O}(10^{-4})$ due to a finite number of samples, the numerical calculation of $P({\cal D})$ has a sharp cutoff at this value near the right edge.}
	\label{fig_tfree}
\end{figure}

\subsection{Origin of the secondary lightcones}
\label{subsec:onedefect}

\begin{figure}
\includegraphics[width=\linewidth]{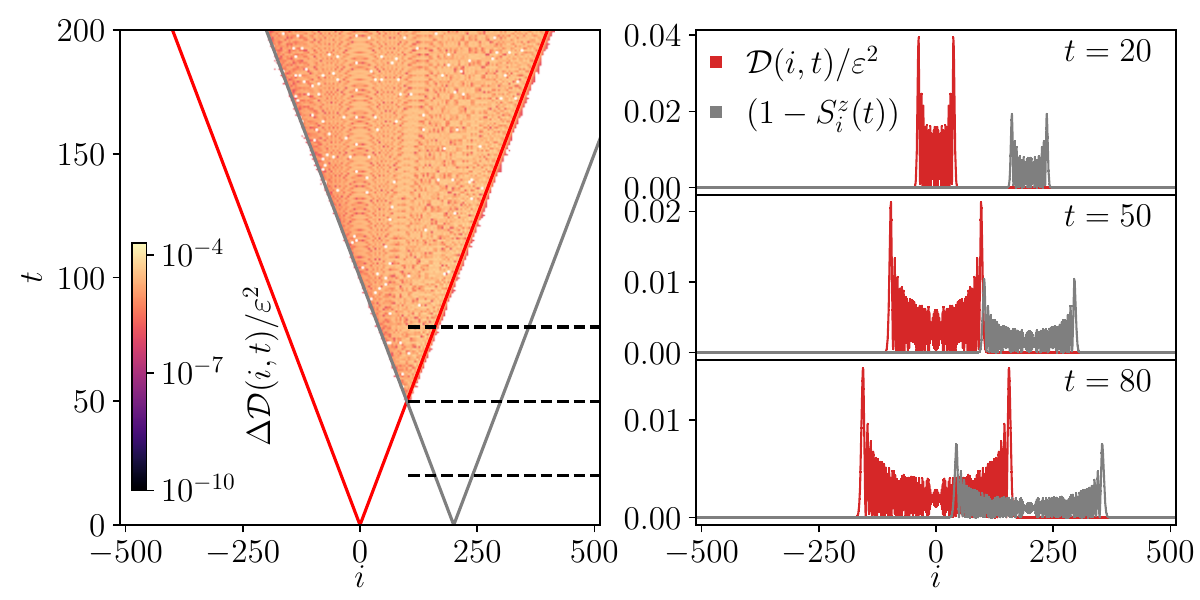}
\caption{Seeding of a secondary lightcone due to a single defect illustrated for the FM case. The left panel shows the difference between the decorrelators $\Delta{\cal D}(i,t)\equiv |{\cal D}_0(i,t)-{\cal D}(i,t)|$ (rescaled by $\varepsilon^2$ for visibility) where ${\cal D}_0(i,t)$ and ${\cal D}(i,t)$ are the decorrelators in the absence of any defects and the presence of a single defect at site $i_d=200$ respectively. The red and grey solid lines denote the boundaries of the primary decorrelator lightcone and the defect dynamics lightcone, respectively. The three dashed horizontal lines denote three different time slices (at $t=20,50,80$) at which ${\cal D}(i,t)$ and the defect profile, $1-S_i^z(t)$ (multiplied by 10 to put it on visible scales) are shown in the right panels, in red and grey respectively. Note that at $t=50$ (right middle panel), the two profiles meet, and that is precisely when the secondary lightcone is seeded at site $i=i_d/2$ and $t_\ast=i_d/(2v_B)$ manifested in $\Delta{\cal D}(i,t)$ taking on a finite value for $t>t_\ast$. Results are for $\theta_d=\pi/10$ and $\phi_d=0$.
}
\label{fig:single-defect-FM}
\end{figure}

The early time regime is immediately followed by a timescale when the defect dynamics start to interact with the primary decorrelator lightcone. We now show explicitly, using numerical and analytical calculations, how they seed the secondary lightcones. To first establish the phenomenology numerically, we simulate the case for a single defect which is placed initially at a site $i_d=d$ while the primary lightcone is seeded at $i=0$.
The primary lightcone, as well as the initially localised defect at site $i_d$, start to spread ballistically with time. Since both the perturbation seeding the lightcone and the defect are completely localised in space, they have overlap with the entire spin-wave spectrum. 
The pilot waves for both the fronts therefore travel with the maximum group velocity $v_B=2J$ and therefore intersect at a time $t_\ast = d/2v_B$.
For $t<t_\ast$, the decorrelator is unaffected by the defect and follows, for example, for the FM, the free Bessel function behaviour, denoted by ${\cal D}_0(i,t)$~\cite{bilitewski2021classical}.  
By contrast, at $t=t_\ast$, the defect-front hits the decorrelator-front at site $i=d/2$ and seeds a secondary lightcone there. 
This is shown clearly in the data in Fig.~\ref{fig:single-defect-FM} where the difference between the free decorrelator and the decorrelator in the presence of a defect, $\Delta{\cal D}(i,t)\equiv |{\cal D}_0(i,t)-{\cal D}(i,t)|$ itself takes the form of a lightcone seeded at $i=d/2$ and $t=t_\ast$.
The results in Fig.~\ref{fig:single-defect-FM} establish numerically the phenomenology that indeed the scattering of the defect dynamics off the primary decorrelator lightcone seeds secondary lightcones.

We now show that the seeding can be understood analytically for the single defect (see  Appendix~\ref{app:SingleDefectAnalytics} for further details).
The starting point for this analysis is the equation for the difference in the spin configurations between the two copies in Eq.~\ref{eq:diffNLin_main}.
The first term on the right-hand side in Eq.~\ref{eq:diffNLin_main} yields the free (Bessel function for FM) solution, ${\cal D}_0(i,t)$.
The presence of a defect, effected by ${\mb{L}_d}\neq 0$, leads to corrections to the free behaviour which can be expanded in orders of $|\mb{L}_d|=\sin\theta_d$.

It is more convenient to analyse these equations in momentum space, where we denote by $\dS_{k}(t)$ the difference between the two copies at momentum mode $k$.
Defining $\bs{\epsilon}_k \equiv \dS_{k}(0)$ as the initial perturbation in $k$-space, the solution  can be split as $\dS_{k}(t) = \dS_k^{(0)}(t) + \dS_k^\mr{int}(t)$.
The former, given by 
\begin{align}
    \dS_k^{(0)}(t) =  \eta_k(\cos(\gamma_k t + \phi_k) \hat{e}_{x} + \sin(\gamma_k t + \phi_k) \hat{e}_{y})  + \epsilon_z \zcap,
    \label{eq_freesol}
\end{align}
is the free solution with $\gamma_k=2J(\cos k-1)$, $\eta_k = (\epsilon_{k,x}^2 + \epsilon_{k,y}^2)^{1/2}$, and $\phi_k = \tan^{-1}(\epsilon_{k,y}/\epsilon_{k,x})$.
The second term, $\dS_k^\mr{int}(t)$, is the correction to the free spin dynamics due to the defect.

With these definitions, the correction to the decorrelator up to leading order is given by
\begin{align}
    \Delta\mathcal{D}^{(1)}&(j, t) = \nonumber \\ &\int_{-\pi}^{\pi} \frac{\d k_1}{2\pi}\frac{\d k_2}{2\pi} e^{-\io (k_1 + k_2) j}\, \big(\dS^{(0)}_{k_1}(t) \cdot \dS^{\mr{int}}_{k_2}(t)\big)\,.
    \label{eq:decorintexp}
\end{align} 
Without loss of generality, we choose the ferromagnetically ordered state to be polarised along the $+\hat{\mb{z}}$ direction.
It is therefore useful to resolve the vectors parallel and perpendicular to the ordering direction as $\dS_k = (\dS_{k, \perp}, \delta S_{k,z})$ and decompose the correction to the decorrelator also similarly,
\begin{multline}
\Delta \mathcal{D}^{(1)}_{\mu}(j,t) = \int_{-\pi}^{\pi} \frac{\d k_1}{2\pi}\frac{\d k_2}{2\pi} e^{-\io (k_1 + k_2) j}
\\\Big(\sum_{\alpha} \delta S_{k_1}^{(0),\alpha}(t)\,\delta S_{k_2}^{\mr{int},\alpha}(t)\Big)\,,
\label{eq_decorrexptrans}
\end{multline} where $\alpha \in (x, y)$ for the perpendicular ($\mu = \perp$) component, and $\alpha = z$ for the parallel ($\mu=\parallel$) component.

As detailed in Appendix~\ref{app:SingleDefectAnalytics}, the parallel (to ordering) component of $\delta{\bf S}$ contains contribution only from linear order in $|\mb{L}_d|$ whereas the perpendicular component contains contributions from both, linear and quadratic, orders in $|\mb{L}_d|$. 
Within the linear spin-wave regime, the first significant correction that captures the onset of secondary lightcone comes from the ${O}(|\mb{L}_d|^2)$ terms in the perpendicular component of the correction.
Systematically including terms order by order, the correction to the perpendicular component of the decorrelator can be written in integral form
\begin{align}
\frac{\Delta\mathcal{D}^{(1)}_{\perp}(j, t)}{\eta^2} = \frac{|\mb{L}_d|^2}{2}\,\mathcal{J}_j(2t)\int_0^t \d t_1 \sum_{m = 0}^{N-1}\bigg[{\cal J}_m(2t_1)\times \nonumber\\
(W_{m,+}(t,t_1)-W_{m,-}(t,t_1))\bigg]\,,\label{eq:FM-single-defect-final}
\end{align}
where
\begin{align}
W_{m,\pm}(t,t_1) = {\cal J}^2_{|m-(i_d\pm 1)|}(2t_1){\cal J}_{|j-(m\mp 1)}(2(t-t_1))\,.
\label{eq:W-func}
\end{align}
and $\eta = (\epsilon_x^2 + \epsilon_y^2)^{1/2}$. While the final result, given in terms of Eq.~\ref{eq:FM-single-defect-final} and Eq.~\ref{eq:W-func}, is somewhat unwieldy, the two essential takeaways are clearer.
First, the convolution of the Bessel functions is such that $\Delta\mathcal{D}^{(1)}_{\perp}(j, t)$ becomes non-negligible only after the decorrelator-front and defect-front hit each other and $j$ lies inside the so seeded secondary lightcone. 
Second, the correction at site $j$ and time $t$, in fact, has a non-local memory kernel.
As such, the correction arises since the defect lies initially at a finite distance from the seeding of the decorrelator.
In Appendix~\ref{app:SingleDefectAnalytics}, we compare the above analytical results with numerical simulations of the perpendicular component of the linearised decorrelator and find these results to be in excellent agreement with each other. A more detailed comparison is shown in Fig. \ref{fig:D1_Exact3}. This concludes our analytical understanding of the origin of the secondary lightcones.

\subsection{Intermediate-time scarred regime} 

The secondary lightcones, whose origins were explained in the above section, and their interactions are the dominant features in the intermediate-time scarred regime, which follows the early-time regime; this manifests itself in the streaky patterns in the heatmap in Fig.~\ref{fig:lc_scarred}.
The statistics of ${\cal D}(i,t)$ in this regime are akin to those of `strong multifractal' distributions~\cite{RevModPhys.80.1355}. 
Physically, this means that there are a few sites where ${\cal D}(i,t)$ is anomalously large, relative to the background of mostly small ${\cal D}(i,t)$. The former corresponds to the set of sites that are within the secondary lightcones seeded by the defects, whereas the latter corresponds to the set of sites where ${\cal D}(i,t)$ continues to take small values characteristic of the bulk of the free solutions. 
This feature leaves its imprint on the distribution of ${\cal D}(i,t)$ over sites $i$ as a power-law distribution, as shown in Fig.~\ref{fig:DDist} (left panel).
More specifically, at timescales where the plateau in $g(t)$ is not fully developed, the distribution $P({\cal D})$ falls off faster than a power law, indicating the overwhelmingly large fraction of sites where ${{\cal D}}\ll 1$ and only a few sites with relatively large ${\cal D}$. 
This is also reflected in the cumulative density function (CDF), which reveals the small span of the magnitudes of $\mathcal{D}$. 

\begin{figure*}
	\centering
	\includegraphics[width=0.8\linewidth]{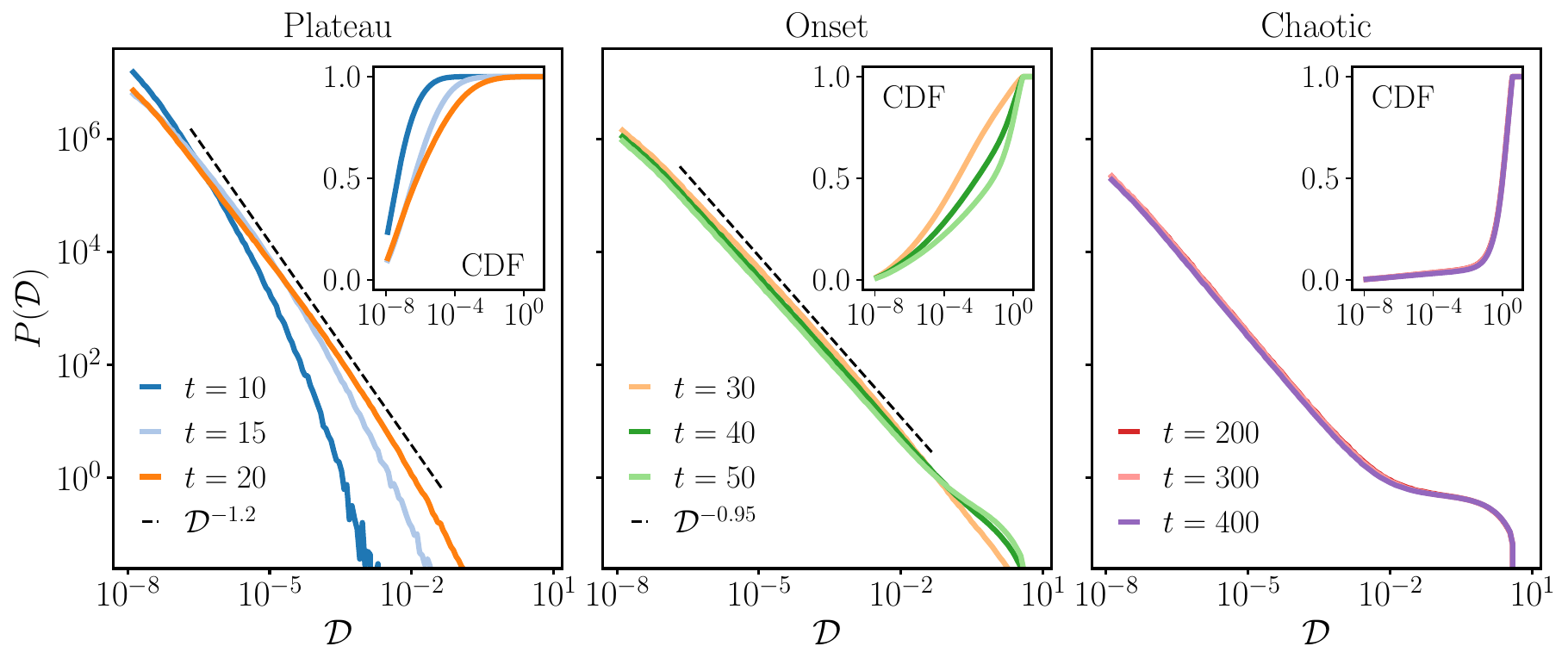}
		\caption{The normalised probability density function ${P}({\cal D})$ of the decorrelator values for defect density $\rho_d=0.5$ over space (and 10240 initial configurations) inside the primary lightcone of an $L = 2048$ lattice with FM coupling. The three panels correspond to three temporal regimes, namely the scarred regime with a plateau in $g(t)$ (left), the regime of onset of chaos (middle), and the chaotic regime (right). The black dashed line in the first two panels denotes the power law decay of the distribution. In the insets, we show the cumulative distribution function (CDF) as a function of decorrelator values ${\cal D}$ for the same time slices.} 
	\label{fig:DDist}
\end{figure*}

To understand the behaviour of $g(t)$ where only a few sites, say $n_s$, out of the total number of sites inside the lightcone, $2v_B t$ at any time $t$ contributes, let us consider the extreme case that the value of the decorrelator on these rare sites is $\mathcal{D}_s$ while that at the rest $n_p=2v_B t-n_s$ sites is $\mathcal{D}_p\sim \varepsilon^2$  where $\mathcal{D}_s/\mathcal{D}_p> 1$ but $\mathcal{D}_s \ll 1$, {\it i.e.}, far away from saturation. It is useful to write $t=t_{\rm free}+\tau$ such that $n_s(\tau=0)=0$. Within this approximation
\begin{align}
    g(t)\sim \frac{\mathcal{D}_s^2 n_s+{\cal D}_p^2n_p}{(\mathcal{D}_s n_s+\mathcal{D}_p n_p)^2}=\frac{1}{2v_B t_{\rm free}}\frac{1+\frac{\mathcal{D}_s+\mathcal{D}_p}{\mathcal{D}_p}\alpha+\frac{\tau}{t_{\rm free}}}{\left[1+\alpha+\frac{\tau}{t_{\rm free}}\right]^2}
    \label{eq_gtanalysis}
\end{align}
where $\alpha=\frac{\mathcal{D}_s-\mathcal{D}_p}{\mathcal{D}_p}\frac{n_s}{2v_Bt_{\rm free}}$. At the beginning of the plateau $\tau/t_{\rm free}\ll 1$ and hence can be neglected. Further, if $n_s<\varepsilon^2 2v_Bt_{\rm free}/2\mathcal{D}_s$, we have 
\begin{align}
    g(t)\sim \frac{1}{4v^2_Bt^2_{\rm free}}\frac{\mathcal{D}^2_s n_s}{\varepsilon^4}
\end{align}
such that for $n_s\sim \mathcal{O}(1)$,  $g\sim t_{\rm free}^{-2}$.
This explains the quasilocalised or `strong multifractal' profile of the decorrelator inside the lightcone, resulting in a scaling of $g(t)\sim t^{-a}$ with $a\gtrapprox 0$. In Fig.~\ref{fig:gt_med_compare}, the approximate plateau or very weak decay of $g(t)$ with $t$ in this regime provides numerical evidence for this picture. Note that as $t_{\rm free}\propto \rho_d^{-1/2}$, the above picture suggests that the plateau height increases linearly with defect density.

However, as the plateau in $g(t)$ develops fully, the distribution $P({\cal D})$ begins to develop a power-law tail. This is the first quantitative indication of the scarred regime being dominated by rare sites with large values of ${\cal D}$, which are the seeds of the higher-order lightcones. In this case, the above conclusion continues to hold -- and hence the plateau -- where $n_s$ now applies to the number of sites with the largest magnitude of the decorrelator.

\subsection{Onset of chaos}
The scarred regime gives way to fully-developed chaos via a proliferation of the higher-order lightcones and scatterings between them, over a rather short time window. 
This is evident in Fig.~\ref{fig:gt_med_compare} where the plateau or slow decay of $g(t)$ rapidly falls off and catches up to the $1/t$ behaviour characteristic of the chaotic regime. 
This rapid fall-off of $g(t)$ provides insight into the mechanism underpinning the onset of chaos from the scarred regime, as we discuss next.

The distribution of ${\cal D}(i,t)$ during this onset of chaos continues to be fat-tailed, described by a power-law, $P({\cal D})\sim {\cal D}^{-1}$ as evidenced by the data in Fig.~\ref{fig:DDist}(middle). The corresponding CDF is marked differently from the left figure for earlier times. Not only does the span of the magnitude of $\mathcal{D}$ cover the entire window till saturation, but also the weights of the high magnitude locations are sizable.
This suggests that the decorrelator is dominated by `hot spots' -- rare sites where it is anomalously large compared to the background -- as is evident from the increased support of $P(\mathcal{D})$ shown in the figure. The anomalously large value of the decorrelator at the hotspots possibly owe their origin to the spin-wave scatterings mediated by the non-linearities which become active almost simultaneously with the overlap of the secondary lightcones seeding the tertiary and higher generation ones.

Heuristically, the decorrelator profile can therefore be modelled as a sum over Dirac-delta functions centred on those rare sites, such that 
${\cal D}(i,t) \propto \sum_{i_{\rm ano}(t)}\delta_{i, i_{\rm ano}(t)}$, where $\{i_{\rm ano}(t)\}$ denotes the set of sites on which the decorrelator is anomalously large at time $t$.
Such a model for the decorrelator would imply that 
\begin{equation}
\sum_i {\cal D}^2(i,t) \propto \sum_i {\cal D}(i,t)\,.
\label{eq:D2=D1}
\end{equation}
The data in Fig.~\ref{fig:AtKtgt}, which shows that, indeed, the above conjecture is satisfied to an excellent approximation over the timescales of the onset of chaos, thereby providing credence to the heuristic model. 
Using Eq.~\ref{eq:D2=D1} in the definition of $g(t)$ in Eq.~\ref{eq:gt_main}, we have
\begin{equation}
    g(t)\approx \frac{1}{\sum_i {\cal D}(i,t)}\,.\label{eq:gt-1/At}
\end{equation}

Alternatively, turning back to Eq. \ref{eq_gtanalysis}, now $\alpha$ dominates over the other terms and hence we have
\begin{align}
    g(t)\sim \frac{1}{n_s}.
\end{align}
Since $\sum \mathcal{D}(i,t)\sim (\mathcal{D}_s n_s+\mathcal{D}_p n_p)\sim n_s$ in this region, the above equation is consistent with Eq. \ref{eq:gt-1/At} and the numerical observation in Fig. \ref{fig:AtKtgt}.

To obtain $g(t)$ in this region, we therefore have to estimate the number of such hotspots. Assuming that the scattering between each pair of higher-order lightcones generates a hotspot at the site of the intersection of their edges, it stands to reason that the number of such hotspots grows exponentially with the number of spin-dynamics lightcones intersecting the primary decorrelator lightcone~\footnote{This is analogous to the picture that in a random branching process, such as a tree, the number of branches at generation $t$ is exponentially large in $t$}. Since the latter is statistically expected to be $O(\rho_d t)$ at time $t$, we have $g(t)\sim e^{-O(\rho_d t)}$, which provides a possible explanation for the rapid fall-off in the data for $g(t)$ between the scarred and chaotic temporal regimes.
This concludes a heuristic picture of an avalanche-like scenario as a possible mechanism underpinning the onset of chaos following the scarred regime. 

\begin{figure}
	\centering
	\includegraphics[width=0.87\linewidth]{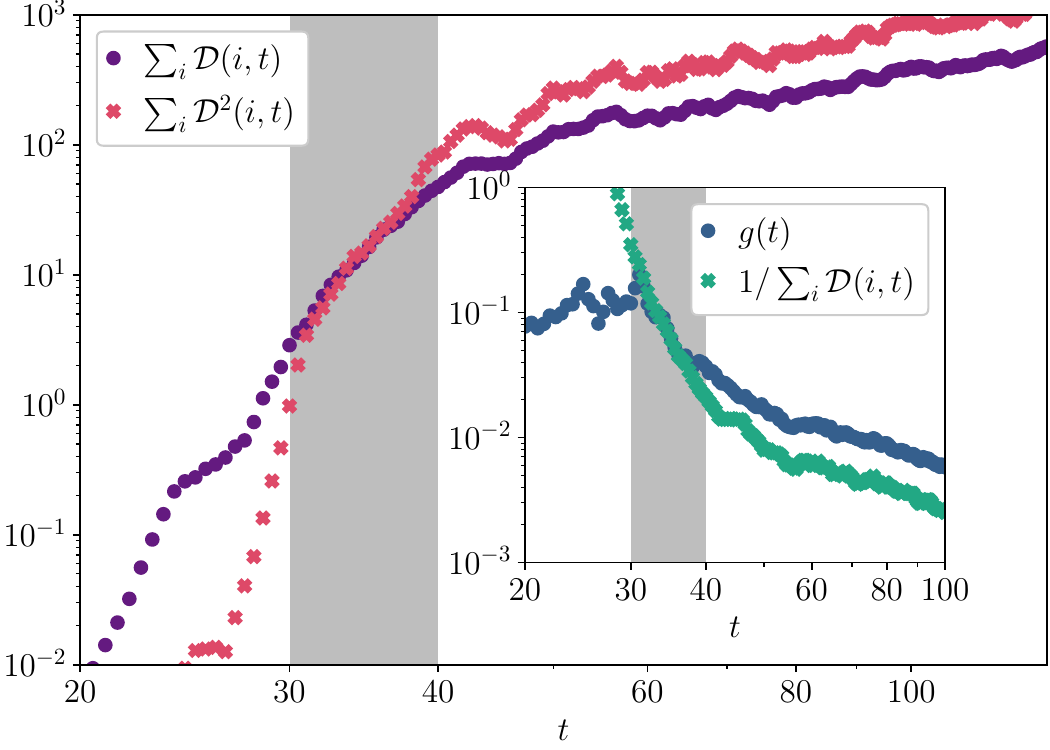}
		\caption{The main panel shows the comparison between $\sum_i{\cal D}(i,t)$ and $\sum_i{\cal D}^2(i,t)$ for a single initial configuration. The data corresponds to $L=2048$ and $\rho_d=0.5$. For this defect density, the best estimate for the onset regime (from the data in Fig.~\ref{fig:gt_med_compare}) is $30<t<40$ indicated by the grey shaded regime and indeed, Eq.~\ref{eq:D2=D1} holds to an excellent approximation in this regime. The inset shows the same phenomenon equivalently by comparing $g(t)$ with $1/\sum_i{\cal D} (i,t)$, suggesting the validity of Eq.~\ref{eq:gt-1/At} in the onset regime.} 
	\label{fig:AtKtgt}
\end{figure}

\subsection{Late-time chaotic regime} 
At late times, the proliferation and scattering of higher-order lightcones lead to fully-developed chaos inside the primary lightcone. 
In such a scenario, the decorrelator becomes approximately homogeneous across space (inside the lightcone) and takes on its saturation value, 
${\cal D}_\mr{sat}$ set by the total initial magnetisation, which is in fact conserved by the dynamics. This is revealed in both $P(\mathcal{D})$ and the CDF in Fig. \ref{fig:DDist} (right). 
For an initial state with an average magnetisation density of $\mb{m}_0$, the saturation is given by ${\cal D}_{\rm sat}=2(1-|\mb{m}_0|^2)$.
In the FM, a defect density of $\rho_d$ leads to an average initial magnetisation of $\mb{m}_0= (1-\rho_d)\hat{e}_{z}$
such that ${\cal D}_{\rm sat}=2\rho_d(2-\rho_d)$.
On the other hand, for the AFM, the average initial magnetisation is zero and therefore ${\cal D}_{\rm sat}=2$ independent of the defect density $\rho_d$.

The approximate homogeneity of the decorrelator inside the lightcone reflects itself in the distribution of ${\cal D}(i,t)$ over space as the fat-tailed power-law distribution in the scarred regime, giving way to a distribution which has a significant weight on $O(1)$ values of ${\cal D}$. This manifests itself as the bump-like feature in the distributions at ${\cal O}(1)$ values of ${\cal D}$ as shown in the right panel of Fig.~\ref{fig:DDist}. 
In fact, this is more clearly seen in the cumulative distribution function (inset), which picks up and rapidly approaches 1 only for ${\cal D}> O(1)$ values. This also implies that $g(t)\sim t^{-1}$. 
To understand this, note that the profile of the decorrelator in this regime can be described approximately as ${\cal D}(i,t)={\cal D}_{\rm sat}\Theta(v_Bt-i)$ modulo the broadening of the front of the edges. Using this form in Eq.~\ref{eq:gt_main} yields $g(t)=1/(v_Bt)$ which explains the $t^{-1}$ scaling. 
The numerical results in Fig.~\ref{fig:gt_med_compare} confirm that this is indeed the case, wherein the black dashed line denotes the $t^{-1}$ behaviour.

A point to note here is that the distributions in Fig.~\ref{fig:DDist} are those of the decorrelator as defined in Eq.~\ref{eq:DecorrMain}, which is bounded from above by 4. Hence, in this saturated, chaotic regime, the mean of the decorrelator is of the same order as its fluctuations (both $O(1)$), resulting in the tail of the distribution at very small values of ${\cal D}$ as well, albeit with a very small weight. 
The distribution of the linearised decorrelator would show a rather sharp peak at a value which grows exponentially with $t$, indicating the relative homogeneity of the decorrelator inside the lightcone. This exponential growth of the averaged linearised decorrelator is evidenced in Fig.~\ref{fig_tfree}(left).

\section{Characteristics of late-time chaos}
\label{sec_lypvb}

\subsection{Defect-density dependence of chaos characteristics}

The late-time chaos in the system can be minimally characterised by two quantities, the Lyapunov exponent (defined in Eq.~\ref{eq:vdle-def}) and the butterfly velocity $v_B$. 
It is therefore natural to study the dependence of these characteristics as a function of $\rho_d$, and understand the correspondences with those as a function of temperature in thermal ensembles~\cite{bilitewski2021classical}.

These quantities can be extracted by studying the linearised decorrelator, $\widetilde{\cal D}(i,t)$ as a function of $t$ along various velocity rays $v=i/t$.
For instance, in Fig.~\ref{fig:lam0} we show the results for $\ln(\braket{\widetilde{\cal{D}}(i=0,t)})/2t$ as a function of $t$ for different values of $\rho_d$.
Note that, at late times, the traces seemingly flatten out and become independent of $t$ such that the saturation value is a faithful proxy for the value in the limit of $t\to\infty$ and hence the value of $\lambda_L(v=0)$. 
The data therein also suggests that $\lambda_L(v=0)$ grows with $\rho_d$ and saturates at larger values of $\rho_d$.
This is made quantitative in Fig.~\ref{fig:LE-vb-rhod}(left) where we show a plot $\lambda_L(v=0)$ as a function of $\rho_d$.
The results show that $\lambda_L(v=0)$ initially grows with $\rho_d$ at small $\rho_d$ before eventually saturating to a $\rho_d$-independent value. 

This can be straightforwardly understood as, at large $\rho_d$, the patches of (quasi) long-range order in the initial states are very short, leading to the state mimicking a short-range correlated state, analogous to a high-temperature state effectively. With the local energy scales bounded, the dynamic scales are therefore set solely by $|J|^{-1}$.
On the other hand, at very low $\rho_d$, the patches of (quasi) long-ranged order are quite large, and over these long length- and timescales, the decorrelator behaviour is akin to that of a zero-temperature state. As the dynamics in Eq.~\ref{eq:spin_dynamics} conserve energy, this leads to an effective restriction in the accessible phase-space volume. 
From the point of view of the decorrelator, the restriction is relaxed via the defects scattering off it; it is therefore natural that $\lambda_L$ depends strongly on the defect density in this regime.

\begin{figure}
\includegraphics[width=\linewidth]{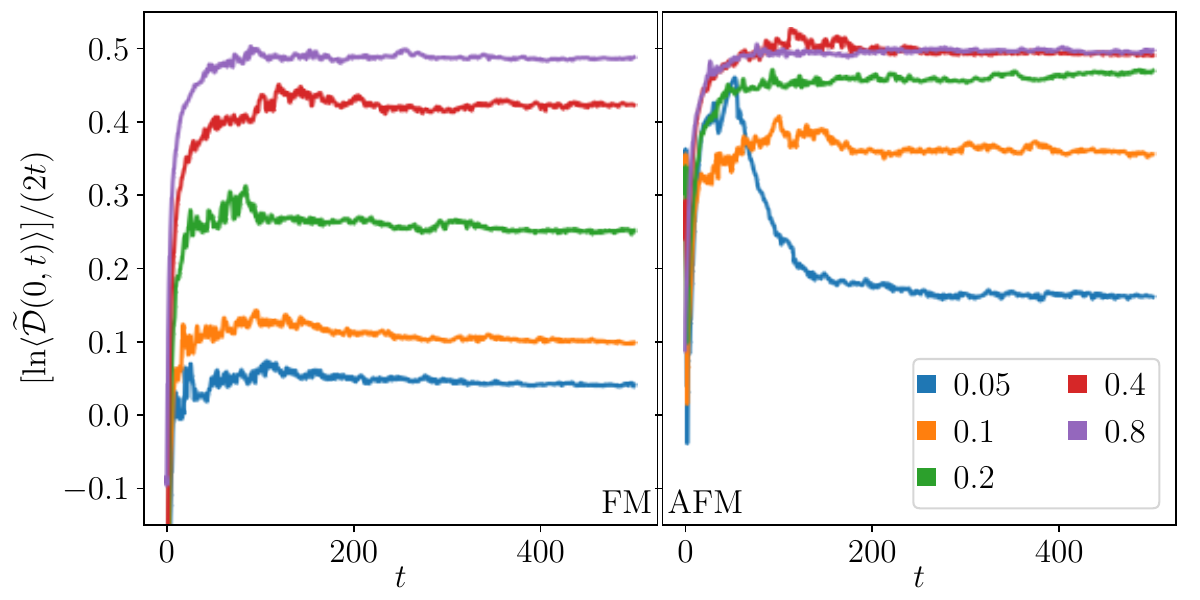}
\caption{The linearised decorrelator $\langle \widetilde{\cal D}(0,t)\rangle$ at site $i=0$ as a function of time, appropriately taken logarithm of and scaled with $2t$. Different colours denote different values of $\rho_d$ as mentioned in the legend. The late time saturation value provides an estimate of the Lyapunov exponent $\lambda_L(v=0)$. 
The left and right panels correspond to the FM and the AFM, respectively. The results are for $L=2048$ and averaged over $\sim 10^4$ initial configurations.}
\label{fig:lam0}
\end{figure}

The analysis used to extract $\lambda_L(v=0)$ can be generalised to arbitrary $v$ by tracking the linearised decorrelator along specific velocity rays at late times and using the expression in Eq.~\ref{eq:vdle-def}. 
The results are shown in Fig.~\ref{fig:vdle} from which two aspects are immediately apparent. First, inside the lightcone, the VDLE grows with increasing $\rho_d$ for all velocities; this is simply the manifestation of the fact that higher $\rho_d$ corresponds effectively to a higher temperature and hence is more chaotic.
Second, the butterfly velocity, indicated by where $\lambda_L(v)$ changes sign, depends very weakly on $\rho_d$. 
This is made quantitative in Fig.~\ref{fig:LE-vb-rhod}(right), which shows that $v_B$ decreases slightly with increasing $\rho_d$ at small values of $\rho_d$ before eventually becoming approximately independent of $\rho_d$.

\begin{figure}[!t]
\includegraphics[width=\linewidth]{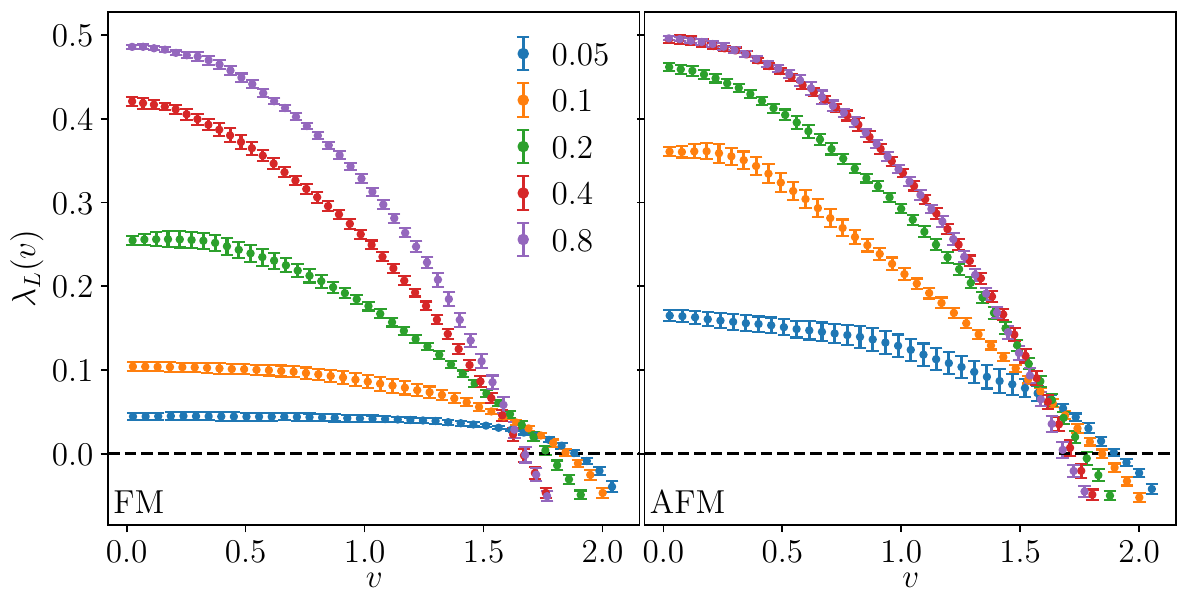}
\caption{The VDLE as a function of $v$ for different defect densities (denoted by the different colours as mentioned in the legend). They are extracted by tracking the saturation (with $t$) value of $[\ln\braket{\widetilde{{\cal D}}(vt,t)}]/2t$ along different velocity rays. The errorbars denote the fluctuations in $[\ln\braket{\widetilde{{\cal D}}(i=vt,t)}]/2t$ over different space-time points $(i,t)$ on the same velocity ray.
The results are for $L=2048$ and averaged over $\sim 10^4$ initial configurations.}
\label{fig:vdle}
\end{figure}

This dependence of $v_B$ on $\rho_d$ can again be understood using arguments similar to those for $\lambda_L$.
At very low $\rho_d$, there are well-defined quasiparticles with rare scatterings between them and also between them and the primary lightcone. Hence, the decorrelator front picks out the maximum group velocity, $\propto |J|$, over the spectrum of excitations and propagates with that.
On increasing $\rho_d$ in this regime, the excitations with lower group velocities but possibly higher density of states also scatter off the primary lightcone; this interplay decreases $v_B$ ever so slightly.
At much larger $\rho_d$, there are no well-defined quasiparticles, but the lightcone continues to be ballistic. In the absence of any other energyscale, the velocity is again set by the coupling constant $|J|$ and is therefore essentially constant with $\rho_d$.

\begin{figure}[!t]
\includegraphics[width=\linewidth]{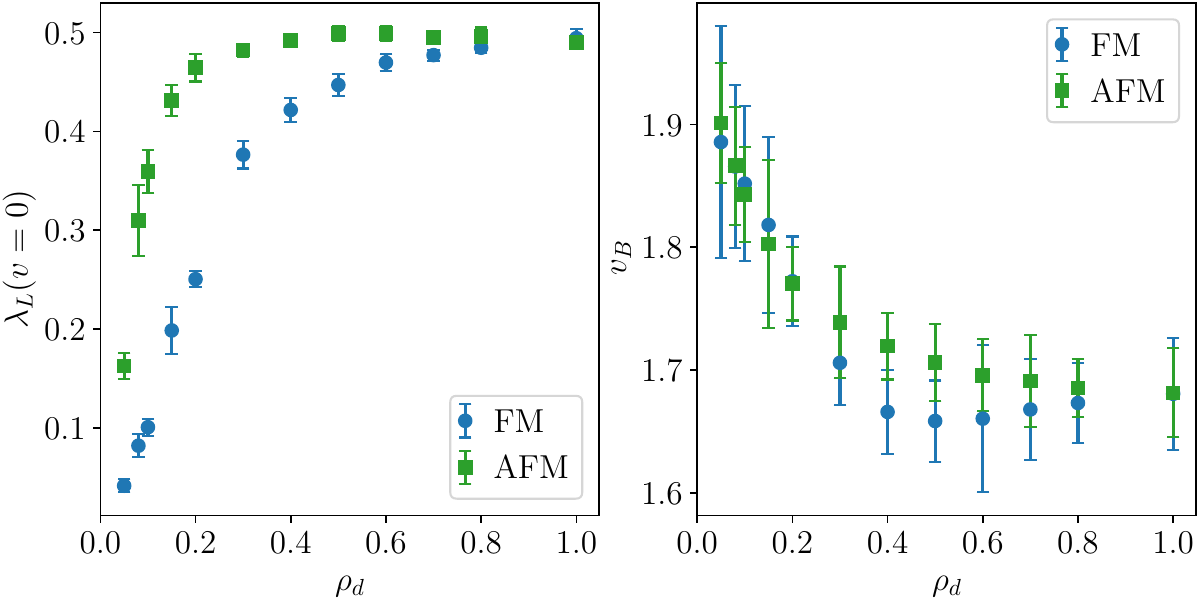}
\caption{Left: The Lyapunov exponent, $\lambda_L(v=0)$, at zero velocity as a function of defect density of $\rho_d$. They are extracted by averaging the traces in Fig.~\ref{fig:lam0} for $t>400$ and the errorbars denote the fluctuations in the time traces. 
Right: The butterfly velocity, $v_B$, as a function of $\rho_d$ extracted by tracking where the VDLE, $\lambda_L(v)$, in Fig.~\ref{fig:vdle} changes sign. The errorbars are obtained by tracking the range of velocities over which the VDLE's fluctuations over different space-time points itself straddles zero.}
\label{fig:LE-vb-rhod}
\end{figure}

\subsection{Sub-ballistic core in the lightcone}
\label{subsec_core}
We finally discuss a notable feature of the global picture: the build-up of a sub-ballistic core in the lightcone.  
To probe this, we introduce a normalised version of the linearised decorrelator as 
\begin{equation}
\widetilde{\cal D}_{\rm norm}(i,t) = \left\langle \frac{{\widetilde{\cal D}}(i,t)}{{\sum_i\widetilde{\cal D}}(i,t)}\right\rangle\, .
\label{eq:D-lin-norm}
\end{equation}
Note that the normalisation is done at the level 
of the decorrelator for each initial configuration, and then the normalised decorrelator is averaged over the realisations.

Fig.~\ref{fig:contours} shows the results for the resulting contours of $\widetilde{\cal D}_{\rm norm}$. The key point is that the contours bend inwards, indicating their sub-ballistic profile.
In the following, we discuss this as a manifestation of the profile of the VDLE with $v$ (see Fig.~\ref{fig:vdle}) and provide a heuristic picture for this in terms of the density of scattering defects.

The linearised decorrelator can be expressed in general in the form,
$
\widetilde{\cal D}(i,t) \sim e^{t\lambda_L(v=i/t)}\,.
$
Taking inspiration from the data in Fig.~\ref{fig:vdle}, we postulate a form of the VDLE as 
\begin{equation}
\lambda_L(v) = \lambda_0\left[1-\left(\frac{|v|}{v_B }\right)^\gamma\right]\,,
\label{eq_vlde}
\end{equation}
where $\lambda_0\equiv \lambda_L(v=0)$ and $\gamma$ appears to decrease from a large value towards 2 with increasing $\rho_d$, evident from the inverted parabolic shape of the VDLE profile.
For this form of the VDLE
\begin{equation}
\widetilde{\cal D}_{\rm norm}(i,t) = \frac{\exp\left[{-\lambda_0 \left(\frac{i}{v_Bt^{1-1/\gamma}}\right)^{\gamma}}\right]}{c_0t^{1-1/\gamma}}\,,
\label{eq:Dnorm-postulate}
\end{equation}
where the constant $c_0$ depends on the parameters $v_B$, $\lambda_0$, and $\gamma$. The contours of $\widetilde{\cal D}(i,t)_{\rm norm}={\cal D}_\ast$ are thus described by 

\begin{equation}
i_\ast(t) \approx \frac{v_B}{\lambda_0^{1/\gamma}}|\ln {\cal D}_\ast|\times t^{1-1/\gamma}\,,
\end{equation}
neglecting a logarithmic in $t$ correction.
Given $\gamma$ appears to be $\ge 2$ in Fig.~\ref{fig:vdle}, the contour $i_\ast(t)$ is sub-ballistic/super-diffusive. It tends towards diffusion, $\gamma \to 2$, for sufficiently large $\rho_d$ when the initial defect ensemble has spin rotation symmetry and hence   $\sum_i\delta{\bf S}_i(t)$ (Eq. \ref{eq:DecorrMain}) is conserved.  Therefore, the average decorrelator, $\propto \left[\delta{\bf S}_i(t)\right]^2$, shows diffusive scaling as discussed in Ref. \cite{bilitewski2021classical}.

As an aside, we note that besides the contours of $\widetilde{\cal D}_{\rm norm}$, one can study the distance $i_f$ from the origin within which a fraction of $f$ of the total mass of decorrelator lives:
\begin{equation}
\sum_{i=-i_f(t)}^{i_f(t)}\widetilde{\cal D}_{\rm norm}(i,t) = f\,.
\end{equation}
Again using the form in Eq.~\ref{eq:Dnorm-postulate}, one obtains
\begin{equation}
i_f(t) = t^{1-1/\gamma}\times \frac{v_B}{\lambda_0^{1/\gamma}}
G^{-1}\left(\frac{\lambda_0^{1/\gamma}}{v_B}c_0 f\right)\,,
\end{equation}
where $G(x) = \int_{-x}^x dz e^{-z^\gamma}$.
$i_f(t)$ thus has the same sub-ballistic/super-diffusive scaling as $i_\ast(t)$, reflecting the dependence of the VDLE on  $v$.

\begin{figure}[!t]
\includegraphics[width=\linewidth]{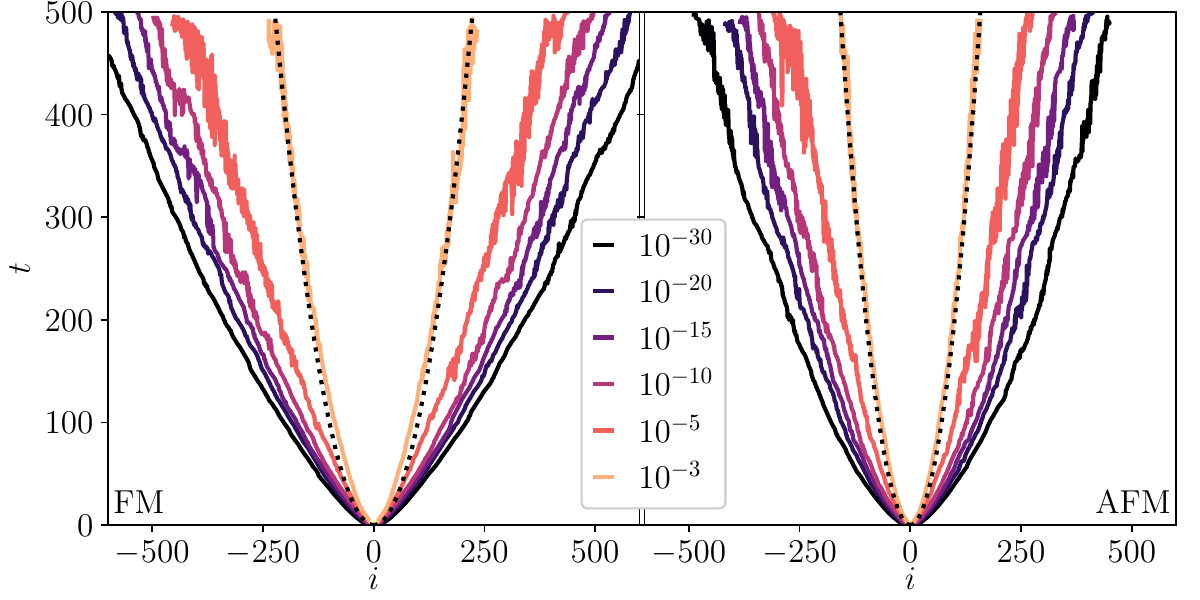}
\caption{Contours of the linearised, normalised decorrelator, $\widetilde{\cal D}_{\rm norm}$, Eq.~\ref{eq:D-lin-norm}, for the FM/AFM (left/right). Different colours correspond to different values of $\widetilde{\cal D}_{\rm norm}$ as mentioned in the legend. All contours bend inwards, indicating a build-up of weight in a core that spreads sub-ballistically. The dotted line is a fit of the highest density curve to a `diffusive' power law, $\sqrt{t}$. 
Data are for $L=2048$, $\rho_d=0.3$, averaged over $\sim 10^4$ initial configurations.}
\label{fig:contours}
\end{figure}

The main obstacle for modelling the (possibly universal) aspects of the normalised decorrelator, $\widetilde{\cal D}_{\rm norm}$, Eq.~\ref{eq:D-lin-norm}, is the non-locality inherent in its normalisation process: growth at one point in space suppresses the value of the decorrelator at an arbitrary distance. This in turn suggests that for a sufficiently uniform/smooth spatial dependence of the normalised decorrelator, a heuristic treatment may be possible by approximating it as a {\it locally} rather than just {\it globally} conserved quantity.  This is most likely the case for high defect density, where isolated scars of the decorrelator are rare on account of their relatively dense seeding by the defects.

We note that Eq.~\ref{eq:Dnorm-postulate} has the appearance of a scaling solution of the form
\begin{equation}
\widetilde{\cal D}_{\rm norm}(i,t) = \frac{1}{t^{1-1/\gamma}}g\left(\frac{i}{t^{1-1/\gamma}}\right)\,.
\end{equation}
One possible equation of motion that admits such a scaling solution (for a class of scaling functions, $g(y)$) is given by
\begin{equation}
\partial_t n(x,t) = \partial_x [n^{-\mu}(x,t)\partial_x n(x,t)]\,,
\label{eq:dens-diff}
\end{equation}
with $\mu = (\gamma-2)/(\gamma-1)$.

The form of the equation in Eq.~\ref{eq:dens-diff} can {\it post facto} be rationalised as it (i) manifestly conserves the total integrated density, $\int dx~ n(x,t)={\rm constant}$, and (ii) is of the form of a diffusion equation where the diffusion constant decays with the density as $\sim n^{-\mu}$~\cite{tikhomirov1991study}. 
 A possible motivation for the latter is the following. 
The scattering events have two effects. 
First, they amplify the decorrelator locally, leading to larger-than-average $n(x,t)$ at those sites.
Second, they disrupt the free ballistic propagation of the decorrelator by creating regimes of larger decorrelators in the interior of the lightcone, thereby adding to the inner core. This suggests that a larger $n(x,t)$ locally disrupts the spread of the decorrelator more, whence a diffusion constant that decays with density in Eq.~\ref{eq:dens-diff}.

With the initial condition $n(x,t=0)=\delta(x)$, Eq.~\ref{eq:dens-diff} has a solution of the form
\begin{equation}
n(x,t)\propto \left[\frac{t}{x^2 + c_1 t^{2/(2-\mu)}}\right]^{1/\mu}\,.
\label{eq:diff-soln}
\end{equation}
The contours of $n(x,t)$ at $n_\ast\ll 1$ then follow
\begin{equation}
x_\ast(t) \sim t^{1/2}n_\ast^{-\mu/2}\,,
\end{equation}
up to subleading, $\mu$-dependent powers of $t$.
This recovers the diffusive spreading of the contours, characteristic of high $\rho_d$. 
As a matter of numerical demonstration, the black dashed curves in Fig.~\ref{fig:contours} denote parabolic fits to the contours of $\widetilde{\cal D}_{\rm norm}$.

\section{Summary and outlook}
\label{sec_summary}

The search for a generic mechanism underpinning the crossover from low-temperature integrable-like behaviour to fully-developed chaos at late times in many-body systems constituted the main motivation for this work. 
The question stems from the fact that a wide class of condensed matter systems, with and without spontaneously broken symmetry, allow for low-energy/temperature descriptions in terms of weakly interacting quasiparticles. The leading-order harmonic Hamiltonian governing the dynamics of these low-energy modes amounts to an integrable description. Yet, the underlying weak non-linearities ultimately lead to equilibration, enabling a description of the system in terms of kinetic theory and/or hydrodynamics. 
The validity of such a coarse-grained description crucially hinges on the emergence of chaos at microscopic scales, thereby sharpening the question of how near-integrable low-temperature dynamics crosses over to chaotic behaviour at late times.

In this work, we addressed this question by analysing the spatiotemporal structure of a standard chaos diagnostic—the classical decorrelator—in paradigmatic classical many-body systems, namely spin chains with nearest-neighbour interactions. This allowed us to identify distinct stages of the crossover and to elucidate the physical processes governing each regime. At early times, quasiparticle interactions are negligible, and the decorrelator exhibits integrable-like short-time behaviour. At intermediate times, nonlinear scattering of quasiparticles with the primary lightcone of the decorrelator gives rise to a scarred regime characterised by a cascade of secondary light cones. The ensuing overlap of these secondary light cones triggers an avalanche of scattering events, ultimately driving the system towards fully-developed chaos.

These systems, at low temperature when the correlation length diverges, allow for an effective description in terms of long-lived harmonic spin-waves.
At early times, quasiparticle interactions are negligible, and the decorrelator exhibits integrable-like short-time behaviour. At intermediate times, nonlinearities lead to scattering of quasiparticles with the primary light cone of the decorrelator, giving rise to a scarred regime characterised by a cascade of secondary lightcones. The ensuing overlap of these secondary lightcones triggers an avalanche of scattering events, ultimately driving the system towards fully-developed chaos.

As a tractable, yet faithful, proxy for the thermal ensemble, we consider a suitably defined ensemble -- the defect ensemble. By controlling the density of excitations in a suitably defined ensemble of initial conditions, the defect ensemble,
we mimic the thermal ensemble at different temperatures.
The defect ensemble used in this work, though different from the thermal ensemble studied in Ref. \cite{bilitewski2021classical}, shares a common phenomenology regarding the onset of chaos. Indeed, at higher density of defects and small deviations from the direction of ordering, the defect ensemble essentially mimics the continuous fluctuations in the thermal ensemble.

The various stages leading to chaos identified in this work would appear to be generic and should hence be present in a wide variety of situations where the system allows for weakly interacting, well-defined harmonic modes (such as spin-waves). The late-time chaotic regime can then be described within a kinetic theory of the weakly interacting harmonic modes/quasiparticles. Thus, the various stages of crossover to chaos form the basis of the applicability of kinetic theory in such systems based on the ergodic behaviour of such weakly interacting harmonic modes or quasiparticles.

While our analysis focused on a classical many-body system, an immediate and natural question concerns the fate of the mechanism we have uncovered in quantum many-body settings. None of the salient features of this mechanism are intrinsically classical, suggesting that the emergence of quantum many-body chaos may likewise be governed by similar physical processes. In particular, the central ingredients, namely, the existence of long-lived quasiparticles (with, however, possibly different amplitudes/densities on account of their quantum statistics) at low energies and weak interactions among them, are also generic to interacting quantum systems.

The usual obstacles to identifying quantum chaos are, of course, ever-present: the finite local Hilbert-space dimension typically leads to a rapid saturation of chaos diagnostics such as OTOCs, precluding the identification of quantities like a Lyapunov exponent in a parametrically long-lived regime outside special cases admitting large-$N$ or semiclassical descriptions. Nonetheless, pinning down signatures of the above mentioned regimes as part of settling the broader question of a generic mechanism governing the emergence of quantum many-body chaos at late times from low-temperature near-integrability clearly merits detailed investigation, both theoretically and, potentially, on near-term quantum simulation platforms.

\acknowledgments
The authors thank B. Doucot, S. Banerjee, A. Kundu, A. Dhar and S. S. Ray for useful discussions. SB and RM thank T. Bilitewski for discussions in the context of previous collaboration on related topics. S. Roy acknowledges support from SERB-DST, Government of India, under Grant No. SRG/2023/000858 and from a Max Planck Partner Group grant between ICTS-TIFR, Bengaluru and MPIPKS, Dresden.
SB acknowledges funding by the Swarna Jayanti fellowship of SERB-DST (India), Grant No. SB/SJF/2021-22/12; DST, Government of India (Nano mission), under Project No. DST/NM/TUE/QM-10/2019 (C)/7 and Max-Planck partner group between ICTS-TIFR, Bengaluru and MPIPKS, Dresden. S. Ruidas, S. Roy and SB acknowledge support of the Department of Atomic Energy, Government of India, under project no. RTI4001.
 This work was in part supported by the Deutsche Forschungsgemeinschaft under grants SFB 1143 (project ID 247310070) and the cluster of excellence ct.qmat (EXC 2147, project-id 390858490).

\appendix

\section{Effective `temperature' of the defect ensemble  \label{appen:ThermalCorrespondence}}

The defect ensemble of initial states discussed in Sec.~\ref{sec:defect-ensemble} is employed as a proxy for a thermal ensemble, with the parameters $(\rho_d,\theta_M)$ controlling the effective temperature, $T_{\rm eff}$, of the corresponding thermal ensemble.
Of course, the defect ensemble at $t=0$ is not equivalent to a thermal ensemble as arbitrary correlation functions in the two need not coincide.
Nevertheless, we can assign a temperature $T_{\rm eff}$ to the defect ensemble from its average energy density $\braket{\mathfrak{e}}_{(\rho_d,\theta_M)}$, and thermalisation at late times would imply that the ensemble of states at late times is indeed the thermal ensemble with temperature $T_{\rm eff}$.
Specifically, for any given $(\rho_d,\theta_M)$ pair, we estimate numerically the average energy density $\braket{\mathfrak{e}}_{(\rho_d,\theta_M)}$ in the defect ensemble and equate it to the thermal value $\braket{\mathfrak{e}}_{T_{\rm eff}}$, and thence solve for $T_{\rm eff}$.

The thermal expectation value can be computed from the partition function of the classical Heisenberg chain, which for a chain with $N+1$ spins with open boundary conditions is given exactly by~\cite{Fisher1964}
\begin{equation}
Z_{N+1} = \left(\frac{\sinh (\beta J)}{\beta J}\right)^{N}\,  
\label{eq:e-dens-therm}
\end{equation}
which is also correct for $N$ spins and periodic boundary conditions up to the leading order for $N\gg 1$.
The partition function yields the average energy density in the thermal ensemble as 
\begin{equation}
\braket{\mathfrak{e}}_{T_{\rm eff}} = T_{\rm eff} - \coth(1/T_{\rm eff})\,.
\end{equation}
On the other hand, to compute $\braket{\mathfrak{e}}_{(\rho_d,\theta_M)}$ numerically, we generate $O(10^4)$ initial states for each $(\rho_d,\theta_M)$ and average the energy density over these states. This is shown in Fig.~\ref{fig:thermal}(left).
The corresponding $T_{\rm eff}$ as a function of $(\rho_d,\theta_M)$ is shown in Fig.~\ref{fig:thermal}(right).
Consistent with expectations, the effective temperature is the lowest (highest) for the smallest (largest) values of $(\rho_d,\theta_M)$.

\begin{figure}
\includegraphics[width=\linewidth]{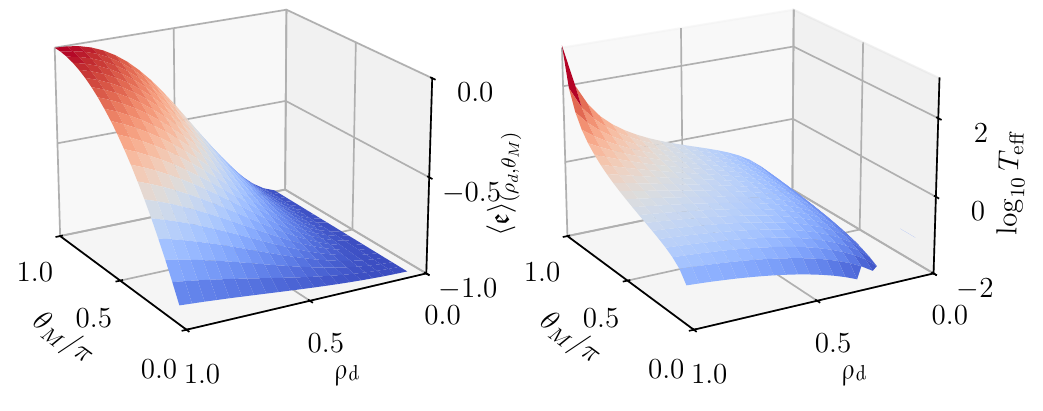}
\caption{Left: The average energy density of the defect ensemble as a function of $(\rho_d,\theta_M)$. The data is obtained by sampling $O(10^4)$ configurations for each $(\rho_d,\theta_M)$ point for $L=2048$. Right: Effective temperature of the defect ensemble as a function of $(\rho_d,\theta_M)$ obtained by equating the data on the left panel to the thermal form in Eq.~\ref{eq:e-dens-therm}. Note that the data is on logarithmic scales.}
\label{fig:thermal}
\end{figure}

We also discuss briefly the dynamical spin correlations in the defect ensemble, defined as
\begin{equation}
\begin{split}
    C_{\perp}(r, t) &= \Braket{\frac{1}{N}\sum_{k=1}^{N}\Big[S_k^x(0)S_{k+r}^x(t) + S_k^y(0)S_{k+j}^y(t)\Big]}\,,\\
    C_{\parallel}(r, t) &= \Braket{\frac{1}{N}\sum_{k=1}^{N} S_k^z(0)S_{k+r}^z(t)}- \Braket{\frac{1}{N}\sum_{k=1}^{N} S_k^z(0)}^2\,,
\end{split}
\label{eq:dyncorr}
\end{equation} 
where $C_\perp$ and $C_\parallel$ are the perpendicular (transverse) and parallel (longitudinal) components of the correlation functions, and the average, denoted $\braket{\cdots}$, is over the initial states drawn from the defect ensemble.

The results are shown in Fig.~\ref{fig:dyncorr-def-ens} where we have taken $\theta_M=\pi$.
For $\rho_d=0.5$ (left column), both $C_{\perp}(r,t)$ and $C_{\parallel}(r,t)$ show diffusive scaling as the curves collapse onto a form $C_{\perp/\parallel} = t^{-1/2}f_{\perp/\parallel}(x/t^{1/2})$.
This is consistent with $\rho_d=0.5$ being representative of high-temperature, where the model is chaotic. This, along with the conservation of magnetisation, mandates diffusion of the correlation functions~\cite{gerling1990time}.

For $\rho_d=0.05$ (right column), representative of low temperature, the dynamics of $C_{\perp}(r,t)$ and $C_{\parallel}(r,t)$ are completely different. 
$C_{\perp}(r,t)$ again shows diffusion due to the $U(1)$ conservation in their $x-y$ plane.
On the other hand, $C_{\parallel}(r,t)$ shows ballistic spreading. This is again consistent with $\rho_d=0.05$ corresponding to low temperature, $C_{\parallel}(r,t)$ reflecting the ballistic propagation of the quasiparticles excited by the defect sites.
\begin{figure}
\includegraphics[width=\linewidth]{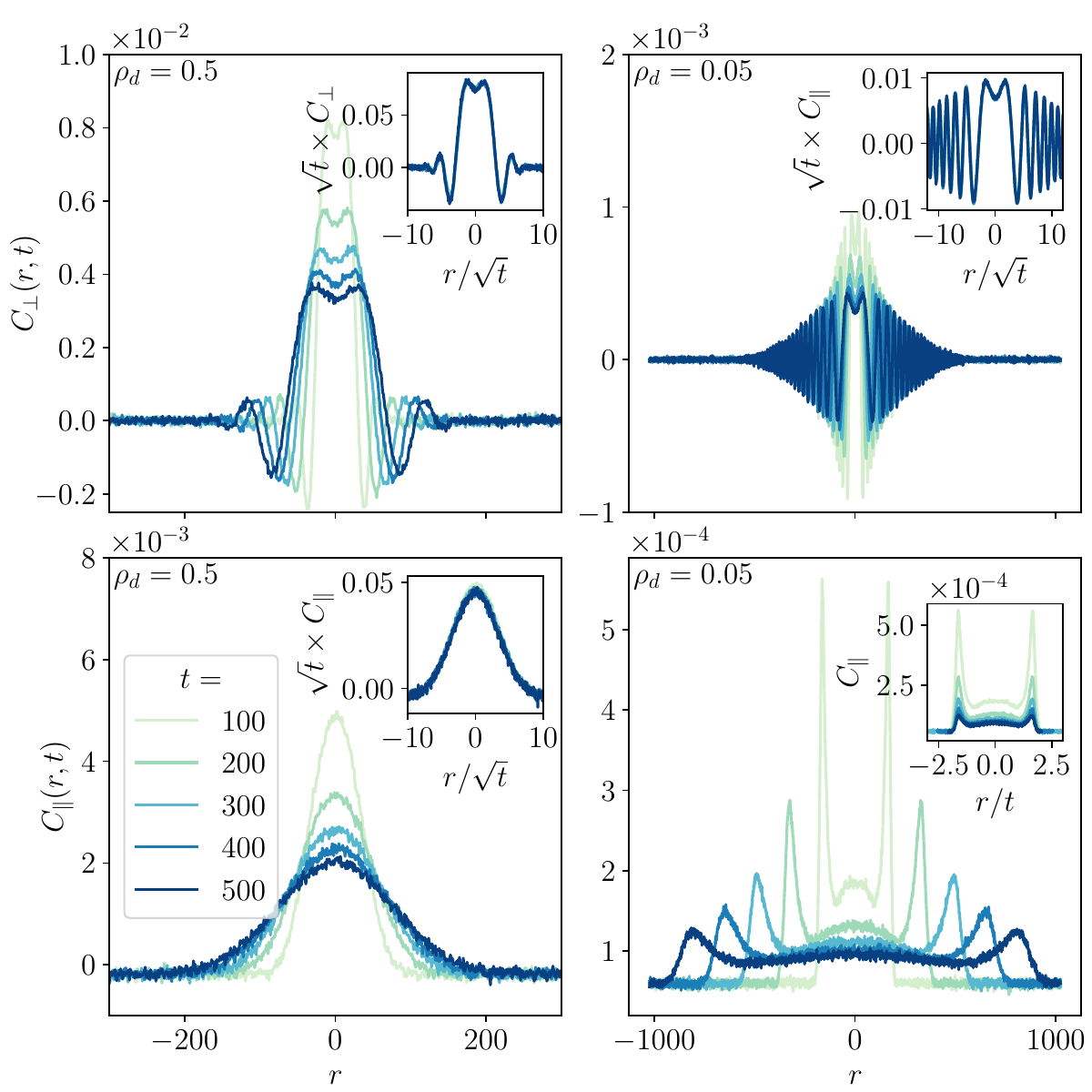}
\caption{Dynamical spin correlations (defined in Eq.~\ref{eq:dyncorr}) in the defect ensemble. The top and bottom rows correspond to the perpendicular and parallel components, respectively. The left and right columns show data for $\rho_d=0.5$ and $0.05$, respectively. The insets show scaling collapses of the data to show the diffusive or ballistic spreading of the correlations. All data for $L=2048$ for a ferromagnetic chain.}
\label{fig:dyncorr-def-ens}
\end{figure}

\section{Correction to the free decorrelator: single defect nonlinear spin-wave calculation \label{app:SingleDefectAnalytics}}

The spin dynamics, as given by Eq.~\ref{eq:spin_dynamics}, can be used to write the equation for the linearised dynamics for the difference, $\dS_i(t) \equiv \mb{S}_{i}^{\mr{a}}(t) - \mb{S}_{i}^{\mr{b}}(t)$, for ferromagnetic interactions as given by Eq. \ref{eq:diffNLin_main}. For an isolated defect at $j = i_d$, considered in Sec. \ref{subsec:onedefect}, we have $\mb{L}_j(0) = \mb{L}_d \, \delta_{j, i_d}$ with $\mb{L}_d = (\sin\theta_d \cos\phi_d, \sin\theta_d\sin\phi_d, 0)$ for a given realisation. Within the linearised dynamics, this spin-wave packet evolves~\cite{bilitewski2021classical} as {given by Eq. \ref{eq:defect-dynamics-linear}.}  From this, the  dynamics of the spin-wave excitation emanating from the defect, $M_j(t) \equiv 1 - S_j^z(t)$, is obtained as \begin{equation}
     M_j(t) =  1- \Big(1 - \big(|\mb{L}_d|\, \mathcal{J}_{|j - i_d|}(2t)\big)^2 \Big)^{1/2}. \label{eq:DefFront_supp}
\end{equation} 
The comparison of the above analytical result and the one obtained from numerics is shown in Fig. \ref{fig:def_bessel} for defect strength $\theta_d = \pi/10$.

\begin{figure}
	\centering
	\includegraphics[width=1.0\linewidth]{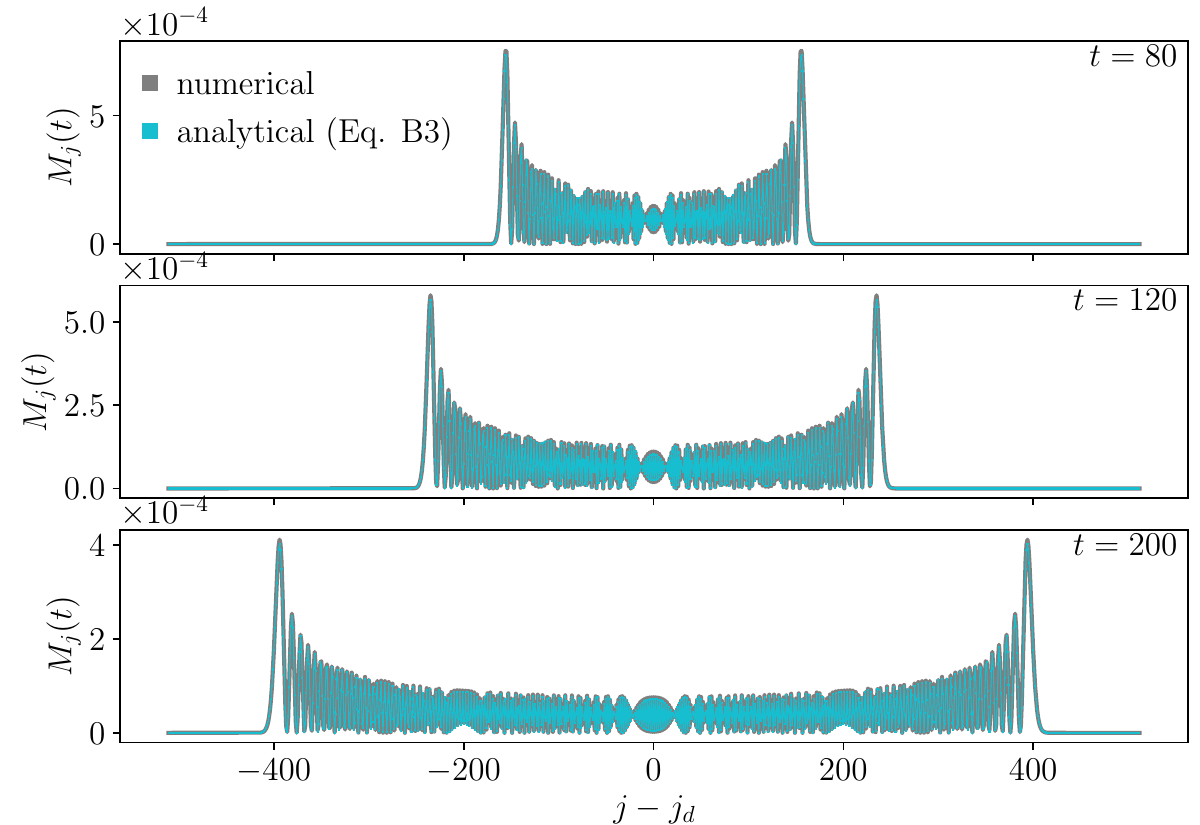}
		\caption{Comparison between the analytical result [Eq.~\ref{eq:DefFront_supp}] and numerical data for the dynamics of the spin-wave excitation from an isolated defect at site $j_d$. Three different time slices at $t = 80, 120$, and 200 are shown. The defect has $\theta_d = 0.1 \pi$ and $\phi_d = \pi/2$ in the background of ordered spins for ferromagnetic interaction.} 
	\label{fig:def_bessel}
\end{figure}

The decorrelator in Eq.~\ref{eq:DecorrMain} can be expressed in Fourier (momentum) space as 
\begin{equation}
    \mathcal{D}(j, t) = \frac{1}{N^2}\sum_{k_1, k_2 \in \mr{BZ}} e^{-\io (k_1 + k_2) r_j}\, \big( \dS_{k_1}(t) \cdot \dS_{k_2}(t) \big)\,, \label{eq:D_onedef}
\end{equation} 
where $k_1, k_2$ are the lattice momenta restricted within the first Brillouin zone $[-\pi, \pi]$ and $\delta{\bf S}_k(t)$ is the Fourier transform of 
$\delta{\mathbf{S}}_i(t)$.
The latter, therefore, satisfies the equation of motion,
\begin{equation}
   \frac{\d\,\dS_k }{\d t}  = \gamma_k \mc{Z} \cdot \dS_k + \frac{1}{N}\sum_{q}  {\cal R}_{k, q} \cdot \dS_{k-q}, \label{eq:diffNLin_k}
\end{equation} where $\mc{Z}$ is the matrix associated with the cross product 
\begin{equation}
    {\cal Z} = \begin{pmatrix}
        0 & -1 & 0 \\
        1 & 0 & 0 \\
        0 & 0 & 0
    \end{pmatrix},
    \end{equation}
and ${\cal R}_{k,q}$ is the time-dependent non-linear spin-wave matrix given by
    \begin{equation}
        {\cal R}_{k,q}(t) = e^{\io q r_d} (\gamma_{k-q} - \gamma_q)\begin{pmatrix} 
     0 & A_q/2 & L_q^y \\
     -A_q/2 & 0 & -L_q^x \\
     -L_q^y & L_q^x & 0
    \end{pmatrix}\,.  \label{eq:Rmat}
\end{equation}
In the above equation, 
\begin{equation}
\mb{L}_q(t) = |\mb{L}_d| \big(\cos(\gamma_q t + \phi_d) \hat{e}_{x} + \sin(\gamma_q t + \phi_d)\hat{e}_{y}\big)\,,
\end{equation} is the Fourier transform of Eq.~\ref{eq:defect-dynamics-linear} and \begin{equation}
    A_q(t) = \int_{-\pi}^\pi \frac{dq'}{2\pi}~~ \mb{L}_{q'} \cdot \mb{L}_{q - q'}\approx |\mb{L}_d|^2\mathcal{J}_{0}\big(4t\sin(q/2)\big)\,,
\end{equation}
where the last expression in the above equation is the leading order in $|\mathbf{L}_d|$.
Using this, to  leading non-linear order, we have 
\begin{widetext}
\begin{equation}
    \delta \mb{S}_k(t) = \Big[ \mc{G}^0_k(t) + \int_{-\pi}^{\pi} \frac{\d q_1}{2\pi} \int_{0}^{t} \d t_1 \,\big(\gamma_{k-q_1} - \gamma_{q_1} \big)\, e^{\io q_1 r_d} \, \mc{G}_{k - q_1}^0(t - t_1) \,{\cal R}_{k, q_1}(t_1) \,\mc{G}^0_{k - q_1}(t_1) + \cdots \Big]\cdot \bs{\epsilon}\,\,,
    \label{eq:skevol}
\end{equation}
\end{widetext}
where $\delta\mb{S}_k(0) = \bs{\epsilon}$ is the infinitesimal initial condition difference that we put to seed the decorrelator at the origin $j = 0$ and $\mc{G}_k^0(t)$ is the Green's function of the linearised spin-wave dynamics \cite{bilitewski2021classical} 
\begin{equation}
    {\cal G}_k^0(t) = \begin{pmatrix}
        \cos(\gamma_k t) & -\sin(\gamma_k t) & 0 \\
        \sin(\gamma_k t) & \cos(\gamma_k t) & 0 \\
        0 & 0 & 1
    \end{pmatrix}\,.
\end{equation}

The defect free evolution is obtained by setting ${\bf L}_d=0$ whence the first term of the LHS of Eq. \ref{eq:skevol} leads to Eq. \ref{eq_freesol}.  Using this in Eq. \ref{eq:D_onedef}, we obtain the free decorrelator~\cite{bilitewski2021classical}
\begin{align}
    \mathcal{D}^0(j,t) = (1 - \eta^2)\delta_{j, 0} + \eta^2 ({\cal J}_{j}(2t))^2 \ .
\end{align}
The deviation from the free propagation, $\Delta \mathcal{D}(j,t) \equiv \mathcal{D}(j,t) - \mathcal{D}^0(j,t)$, is then given by Eq. \ref{eq:decorintexp}. Resolving the vectors $\delta{\bf S}$ into components parallel ($\parallel$) and perpendicular ($\perp$) to the direction of the magnetic ordering, we find that the leading contribution to the decorrelator is from the perpendicular component, which, for the above isolated defect, is given by Eq.~\ref{eq_decorrexptrans}. In Fig. \ref{fig:D1_Exact3}, we plot the analytically calculated $|\Delta \mc{D}^{(1)}_\perp(j,t)|/\eta^2$ from Eq. \eqref{eq_decorrexptrans} and compare it with the full numerical data for $|\mc{D}_\perp(j,t) - \mc{D}^0_{\perp}(j,t)|/\varepsilon^2$, showing excellent agreement.

\begin{figure}
	\centering
	\includegraphics[width=\linewidth]{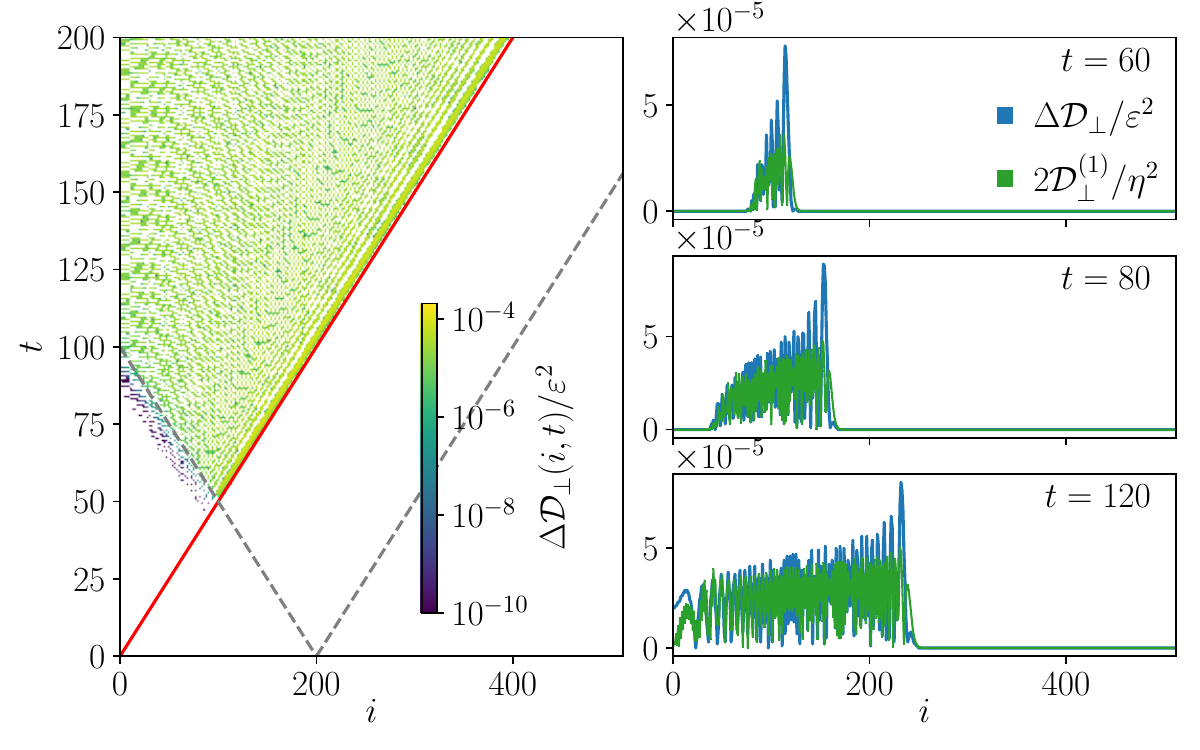}
		\caption{Left: The difference between the perpendicular components of the linearised decorrelators, $\Delta \mc{D}_\perp(i,t) = |\mc{D}_\perp(i, t) - \mc{D}_{\perp}^0(i, t)|$, in the presence of a single defect at site $i_d = 200$.
        Right: Comparison between the analytically computed leading order correction (green lines), $2\mc{D}^{(1)}_\perp(j,t)/\eta^2$ (see Eq.~\ref{eq:FM-single-defect-final}), to the perpendicular component with the numerical results (blue lines), at three different times as mentioned in the panels.} 
	\label{fig:D1_Exact3}
\end{figure}

\bibliography{references}
\end{document}